\documentclass[11pt]{article}

\usepackage{amssymb}
\usepackage{subfig}
\usepackage{color}
\usepackage{epsfig,amssymb,amsfonts,amsmath,graphicx,dsfont,cite,xfrac}
\usepackage{authblk}
\definecolor{mygray}{gray}{0.5}
\usepackage{cite}
\usepackage[colorlinks=true,linkcolor=blue,citecolor=red]{hyperref}

\usepackage{tikz,pgfplots}
\usetikzlibrary{decorations.pathreplacing}
\usetikzlibrary{shapes.multipart}
\usetikzlibrary{positioning}
\usetikzlibrary{matrix}
\usetikzlibrary{shapes,calc,arrows}

\newcommand{\be}{\begin{equation}}
\newcommand{\ee}{\end{equation}}
\newcommand{\bea}{\begin{eqnarray}}
\newcommand{\eea}{\end{eqnarray}}

\title{Coupled system of Dirac fermions with different Fermi velocities via composites of SUSY operators}

\author[${}$]{V. Jakubsk\'y, K. Zelaya  }

\affil[${}$]{\footnotesize Nuclear Physics Institute, Czech Academy of Science, Prague/\v{R}e\v{z}, Czech Republic}

\date{}

\begin{document}

\maketitle

\begin{abstract}
We use the framework of supersymmetric transformations in the construction of coupled systems of Dirac fermions.  Its energy operator is a composite of the generators of the associated superalgebra, and the two coupled Dirac fermions acquire different Fermi velocities. We discuss in detail the peculiar spectral properties of the new system. The emergent phenomena like level crossing or generation of bound states in the continuum (BICs) are illustrated in two explicit examples.
\end{abstract}



\section{Introduction}
Physical systems described by coupled Dirac equations are diverse. In physics of graphene, they appear in description of interlayer scattering of Dirac fermions in bilayer graphene \cite{CastroNeto}, \cite{McCann}, \cite{Falco} qualitatively the same operators are used in the analysis of spin-orbit (Rashba) interaction \cite{Kane}, \cite{Huertas}, \cite{Avsar}. They describe intervalley scattering due to distortions \cite{Altland}, \cite{AndoDisorderScatt}, \cite{Frassdorf}, \cite{AndoCross}, \cite{Koshino}. They appear in description of electron-hole scattering in graphene-superconductor interfaces \cite{Beenaker1}, \cite{Beenaker2}, \cite{Titov}. 

Supersymmetric quantum mechanics provides efficient tools for construction of solvable models. Originally, it was developed for the systems described by Schr\"odinger equation \cite{CooperKhare}, \cite{Junker}. In recent years, it has been implemented for the analysis of physics described by low-dimensional relativistic equations. Supersymmetric transformations for one-dimensional Dirac equation were discussed in \cite{Samsonov1}, \cite{Samsonov2} where the matrix coefficients were represented by $2\times2$ matrices. A more general analysis was conducted in \cite{Schulze1}, see also \cite{Celeita2}, where supersymmetric transformations for $N\times N$ Dirac operators were considered.  Supersymmetry for arbitrary spin relativistic equation was discussed in \cite{Jun20}. Construction of two-dimensional models via supersymmetric transformations was presented recently in \cite{Schulze2}.

In the current article, we will discuss construction of a coupled system of Dirac fermions of non-equivalent Fermi velocity based on supersymmetric (susy) transformation for $2\times2$ Dirac operators. We call the final Hamiltonian, given in terms of $4\times4$ matrices, as the composite of supersymmetric operators. The composite  has spectral properties that differ dramatically from those of the uncoupled supersymmetric Hamiltonian. Contrary to standart framework of susy transformation, the new system is not isospectral with the initial one. Instead, there exists rather nontrivial mapping between the energy levels of the initial and the new model. Let us notice that similar situation was studied in \cite{Jakubsky} where such systems were called spectrally isomorphic.

The article is organized as follows. In sections \ref{sec:DT} and \ref{sec:Ext-Spec}, we review the main features of susy transformation for one-dimensional Dirac operators.  We show that the supersymmetric Hamiltonian together with the supercharges can be used together in construction of the energy operator of an extended coupled system. We discuss solvability of its stationary equation in detail, for real eigenvalues in particular. In section \ref{Examples}, we illustrate the construction on two specific examples based on P\"oschl-Teller exactly solvable model.  The last section is left for discussion.

\section{Supersymmetric operator composites}
\label{sec:DT}
We review the key features of the supersymmetric transformation for one-dimensional stationary equation. We define the extended supersymmetric operators that serve for definition of composite Hamiltonian. Finally, we will discuss two particular scenarios where the system possesses chiral symmetry.

\subsection{Supersymmetric transformation for one-dimensional Dirac equations}
Let us have two one-dimensional stationary Dirac equations
\begin{eqnarray}
&& H\mathbf{\Psi}\equiv\left(-i\sigma_2\partial_{x}+V(x)\right)\mathbf{\Psi}= \lambda\mathbf{\Psi} \, , \label{r1}\\ 
&& \widetilde{H}\widetilde{\mathbf{\Psi}}\equiv\left(-i\sigma_2\partial_{x}+\widetilde{V}(x) \right)\widetilde{\mathbf{\Psi}}=\widetilde{\lambda}\widetilde{\mathbf{\Psi}} \, ,
\label{r2}
\end{eqnarray}
with hermitian operators $H$ and $\widetilde{H}$. The two equations are related by supersymmetric transformation $L$ provided that there holds
\begin{equation}
L H=\widetilde{H}L\,\quad HL^\dagger=L^\dagger \widetilde{H} ,
\label{intert1}
\end{equation}
where the second equality is just conjugation of the first one. The so-called intertwining relations (\ref{intert1}) can simplify the solution of (\ref{r2}) enormously, provided that the solutions of (\ref{r1}) are known. Indeed, when we know solution $\mathbf{\Psi}$ of (\ref{r1}), then the spinor $\widetilde{\mathbf{\Psi}}=L\mathbf{\Psi}$ solves (\ref{r2}) for the same eigenvalue $\widetilde{\lambda}=\lambda$.

If (\ref{r1}) is solvable, we can use its solutions to construct the triplet $H$, $\widetilde{H}$ and $L$ (and $L^\dagger$)  such that the intertwining relations (\ref{intert1}) are satisfied. Following \cite{Samsonov1}, the construction is based on two spinors $u_1$ and $u_2$, $(H-\epsilon_a)u_a=0$, $a=1,2$. They can be used to compose the matrix $U=(u_1,u_2)$ that satisfies
\begin{equation}
\left(-i\sigma_2\partial_{x}+V\right)U=U\Lambda \, , \quad \Lambda=\operatorname{diag}(\epsilon_{1},\epsilon_{2}) \, ,\quad \epsilon_1<\epsilon_2.
\label{eigenU}
\end{equation}
The operators $\widetilde{H}$ and $L$ are then defined as
\begin{equation}
\widetilde{H}=H-i[\sigma_2,U_xU^{-1}],\quad L=\partial_x-U_{x}U^{-1}=U\partial_xU^{-1} \, .
\label{L}
\end{equation}
By construction (see \cite{Samsonov1}), the operators comply with (\ref{intert1}). It has to be mentioned that the construction is formal in the sense that $\widetilde{H}$ can be non-hermitian in general and its potential term $\widetilde{V}$ can have additional singularities when compared to $V$. The spinors $u_1$ and $u_2$, called "seed" solutions, have to be fixed such that $\widetilde{H}$ has the desired properties. We suppose that this is the case in what follows. 

The operator $L^\dagger=-(U^{-1})^\dagger\partial_xU^\dagger$ annihilates two states $\tilde{u}_1$ and $\tilde{u}_2$ such that $(U^{-1})^\dagger=(\tilde{u}_1,\tilde{u}_2)$. The spinors $\tilde{u}_1$ and $\tilde{u}_2$ are called missing states in the literature. They can represent bound states of $\widetilde{H}$ whose analogues are missing in the system described by $H$. It is possible to show, see \cite{Samsonov1}, that the products of $L$ and $L^\dagger$ are polynomials of second order in either $H$ of $\widetilde{H}$,
\begin{equation}\label{LL}
L^\dagger L=(H-\epsilon_1)({H}-\epsilon_2),\quad  LL^\dagger=(\widetilde{H}-\epsilon_1)(\widetilde{H}-\epsilon_2).
\end{equation}
Notice that one or both of $\widetilde{u}_a$ can be squre-integrable despite $u_a$ did not have this property.

\subsection{Composites of the supersymmetric operators}
The link of the triplet $H$, $\widetilde{H}$, and $L$ to the supersymmetry, as we know it in quantum field theory, dwells in the associated algebraic structure. We can define extended operators $\mathbb{H}_0$, $\mathbb{L}_1$ and $\mathbb{L}_2$,
\begin{equation}
\mathbb{H}_0=\left(\begin{array}{cc}H&0\\0&\widetilde{H}\end{array}\right),\quad \mathbb{L}_1=\left(\begin{array}{cc}0&L^\dagger\\L&0\end{array}\right),\quad \mathbb{L}_2=i\Gamma \mathbb{L}_1,\quad \Gamma=\sigma_3\otimes\sigma_0.\label{susyop}
\end{equation}
The extended Hamiltonian $\mathbb{H}_0$ commutes with $\Gamma$ while both $\mathbb{L}_a$ anticommute with this operator. We can consider $\Gamma$ as the grading operator of superalgebra that is established by the following relations,
\begin{equation}
[\mathbb{H}_0,\mathbb{L}_a]=0,\quad \{\mathbb{L}_a,\mathbb{L}_b\}=\delta_{ab}(\mathbb{H}_0-\epsilon_1)(\mathbb{H}_0-\epsilon_2), \quad a,b\in\{1,2\}.
\end{equation}
The intertwinig relations (\ref{intert1}) are encoded in the commutator whereas the relations (\ref{LL}) are reflected by the anticommutator. 
The operator $\mathbb{H}_0$ plays the role of supersymmetric Hamiltonian and $\mathbb{L}_a$ can be considered as the supercharges. In this realization, the "bosonic" sector corresponds to the space spanned by the upper spinors (where the grading operator $\Gamma$ has the eigenvalue equal to one)  and the "fermionic" sector spanned by the lower spinors. 
The superalgebra is nonlinear as the anticommutator of the supercharges is the polynomial of the second order in $\mathbb{H}_0$. It is worth noticing that quantum systems, both relativistic and non-relativistic, possessing nonlinear superalgebra, have attracted considerable attention, see e.g. \cite{NSUSY1}, \cite{NSUSY2}, \cite{NSUSY3} and references therein.

The three operators (\ref{susyop}) do not mutually commute, so that it is not possible to find their common eigenvector. Nevertheless, we can find the common eigenvectors of $\mathbb{H}_0$ and  one of the supercharges. For  $\mathbb{H}_0$ and $\mathbb{L}_1$, they are
\begin{align}
&\Psi^{\pm}=\left(\begin{array}{c}\pm \sqrt{F(\lambda)}\psi\\ L\psi\end{array}\right),\quad F(\lambda)=(\lambda-\epsilon_1)(\lambda-\epsilon_2)\\
&
\mathbb{H}_0\Psi^{\pm}=\lambda\Psi^{\pm},\quad \mathbb{L}_1\Psi^\pm=\pm \sqrt{F(\lambda)}\Psi^{\pm}. \label{bispinor1}
\end{align}
In case of the missing states, we have
\begin{equation}\label{bispinor2}
\Psi_{\tilde{u}_a}=\left(\begin{array}{c}0\\\tilde{u}_a\end{array}\right),\quad \mathbb{H}_0\Psi_{\tilde{u}_a}=\epsilon_a\Psi_{\tilde{u}_a},\quad \mathbb{L}_1\Psi_{\tilde{u}_a}=0 \quad a=1,2.
\end{equation}

Now, we combine the supersymmetric Hamiltonian $\mathbb{H}_0$ together with its integral of motion $\mathbb{L}_1$ into the following \textit{composite of the supersymmetric operators},
\begin{equation}
\mathbb{H}_\alpha=\mathbb{H}_0+\alpha\mathbb{L}_1=
\begin{pmatrix}
H & \alpha L^{\dagger} \\
\alpha L & \widetilde{H}
\end{pmatrix}
\, ,\quad \alpha\in\mathbb{R}.
\label{EXT-H}
\end{equation}
We can immediately find solutions of the associated stationary equation in terms of the bispinors (\ref{bispinor1}) and (\ref{bispinor2}),
\begin{align}
&\mathbb{H}_{\alpha}\Psi^{\pm}=\mathbb{E}^{\pm}(\lambda)\Psi^{\pm},\quad \mathbb{E}^{\pm}(\lambda)=(\lambda\pm\alpha \sqrt{F(\lambda)}),\nonumber\\
&\mathbb{H}_{\alpha}\Psi_{\tilde{u}_a}=\epsilon_a\Psi_{\tilde{u}_a}.\label{Halphastac}
\end{align}
A few comments are in order. The linear combination $\mathbb{H}_0+\alpha_1\mathbb{L}_1+\alpha_2\mathbb{L}_2$ is equivalent to $\mathbb{H}_{\alpha}$ with $\alpha=\alpha_1+i\alpha_2$. As the operators do not commute, the only way to find the corresponding eigenvectors is to fix either $\alpha_2=0$ or $\alpha_1=0$. In the later case, it would be necessary to multiply the upper spinor in $\Psi^{\pm}$ by $i$ to get the corresponding eigenstates.
It is worth noticing that the eigenstates $\Psi^{\pm}$ themselves do not depend on $\alpha$ but their eigenvalues do. The eigenvalues of  the missing states $\Psi_{\tilde{u}_a}$ are constant and remain intact with respect to variations of $\alpha$.

Hamiltonians similar to $\mathbb{H}_\alpha$ were discussed \cite{Jun20}. Therein, this kind of operators had direct physical interpretation in terms of energy operators of relativistic particles of specific spin. In our case, we do not see any direct physical scenario that would be described by this type of operator. Nevertheless, rather straightforward manipulation can turn it into more common form. 
We can define the following matrix
\begin{equation}
\mathcal{U}=\frac{1-i}{2\sqrt{2}}\left(\begin{array}{rrrr}-1&-1&1&1\\1&-1&-1&1\\1&-1&1&-1\\1&1&1&1\end{array}\right).\  (\sigma_0\otimes e^{-i\frac{\kappa}{2}\, \sigma_2}).
\end{equation}
It can be used to transform the Hamiltonian $\mathbb{H}_{\alpha}$ as follows,
\begin{eqnarray}&&\widetilde{\mathbb{H}}_{\alpha}=\mathcal{U}^{-1}\mathbb{H}_\alpha\, \mathcal{U}=-i\left(\begin{array}{cc}(1+\alpha)\sigma_2&0\\0&(1-\alpha)\sigma_2\end{array}\right)\partial_x+\widetilde{\mathbb{V}}.\label{tHalpha}
\end{eqnarray}
This operator can be understood as the Hamiltonian of two coupled Dirac fermions with two different Fermi velocities $1-\alpha$ and $1+\alpha$. In order to keep both values positive, we make the following restriction from now on,
\begin{equation}
\alpha\in[0,1).
\end{equation}
As much as we are not aware of any existing physical realization of (\ref{tHalpha}), it is worth noticing in this context that the Fermi velocity in graphene can be manipulated by perpendicular electric field \cite{Diaz-Fernandez}. Therefore, $\widetilde{\mathbb{H}}_{\alpha}$ could be interpreted as the Hamiltonian of Dirac fermions in bilayer graphene where the Fermi velocity in each layer is manipulated correspondingly. The Dirac fermions in the two layers would interact via the interaction $\widetilde{\mathbb{V}}$.

\subsection{Composites of SUSY operators with chiral symmetry}
\label{sec:susy-chiral}
In general, the potential $\widetilde{\mathbb{V}}=\mathcal{U}^{-1}\mathbb{V}\mathcal{U}$ can have rather complicated form. Let us discuss two particular scenarios where the formulas get considerably simplified due to the chiral symmetry of $\widetilde{H}$, $\{\widetilde{H},\sigma_3\}=0$. In the first case, the initial Hamiltonian $H$ is chiral, 
\begin{equation}
H=-i\sigma_2\partial_x+v_1\sigma_1.\label{Hchiral1}\end{equation}
In order to guarantee chiral symmetry of $\widetilde{H}$, the matrix $U$ is fixed in the following manner \cite{Samsonov1},
\begin{equation} U=\left(\begin{array}{cc}u_{11}&u_{11}\\u_{12}&-u_{12}\end{array}\right),\quad HU=U\Lambda,\quad \Lambda=\left(\begin{array}{cc}\lambda_1&0\\0&-\lambda_1\end{array}\right)
\end{equation}
where we select $(u_{11},u_{12})^T$ such that its components are real functions. The intertwining operator $L$ and the new Hamiltonian $\widetilde{H}$ are then given as
\begin{equation}
L=\partial_x+\left(v_1-\lambda_1\frac{u_{12}}{u_{11}}\right)\sigma_3
,\quad 
\widetilde{H}=-i\sigma_2\partial_x-\left(v_1-\lambda_1\frac{u_{11}^2+u_{12}^2}{u_{11}u_{12}}\right)\sigma_1.
\end{equation}
Let us present explicitly two cases of the composite of operator (\ref{EXT-H}) for $\kappa=0$ and $\kappa=\pi/2$ as the potential term acquires particularly simple form,

\begin{eqnarray}
&&\widetilde{\mathbb{V}}=\begin{cases}
\left(\begin{array}{cccc}m_1&0&V_{13}&V_{14}\\0&-m_1&-V_{14}&-V_{13}\\V_{13}&-V_{14}&m_2&0\\V_{14}&-V_{13}&0&-m_2\end{array}\right),\quad \kappa=0,\\
\\
\left(\begin{array}{cccc}0&-m_1&0&V_{14}-V_{13}\\-m_1&0&-V_{13}-V_{14}&0\\0&-V_{13}-V_{14}&0&-m_2\\V_{14}-V_{13}&0&-m_2&0\end{array}\right),\quad \kappa=\pi/2.\end{cases}\\\label{tVchiral1}
\end{eqnarray}

where
\begin{align}
&m_1=(1+\alpha)(-v_1+V_{13}),\quad m_2=\frac{1-\alpha}{1+\alpha}m_1,&\\
&V_{13}=\frac{\lambda_1(u_{11}^2+u_{12}^2)}{2u_{11}u_{12}},\quad V_{14}=\alpha\lambda_1\frac{u_{11}^2-u_{12}^2}{2u_{12}u_{12}}.&
\end{align}
Therefore, the two Dirac fermions have position dependent masses $m_1$ and $m_2$ for $\kappa=0$, whereas they are massless but in presence of a pseudo-magnetic field for $\kappa=\pi/2$.  

In the second scenario, the initial Hamiltonian $H$ ceases to be chiral symmetric,
\begin{equation}
H=-i\sigma_2\partial_x+v_1\sigma_1+m\sigma_3.
\end{equation}
The matrix $U$ is fixed in this way 
\begin{equation}
U=\left(\begin{array}{cc}0&u_{12}\\u_{21}&u_{22}\end{array}\right),\quad HU=U\Lambda,\quad \Lambda=\left(\begin{array}{cc}-m&0\\0&0\end{array}\right),
\label{U2}
\end{equation}
where $u_{12}$, $u_{21}$, and $u_{22}$ are real functions. Then the intertwining operator and the new Hamiltonian are explicitly
\begin{equation}
L=\partial_x+\left(\begin{array}{cc}v_1-m\frac{\,u_{22}}{u_{12}}&0\\-m&-v_{1}\end{array}\right),\quad
\widetilde{H}=-i\sigma_2\partial_x-\left(v_1-m\frac{u_{22}}{u_{12}}\right)\sigma_1.
\label{L2}
\end{equation}
Defined in terms of these operators, the composite operator (\ref{tHalpha}) can be calculated in straightforward manner. Instead of presenting its general form, we prefer to fix $\kappa=\pi/2$. Then we get
\begin{eqnarray}
&&\widetilde{\mathbb{H}}_{\alpha}=-i\left(\begin{array}{cc}(1+\alpha)\sigma_2&0\\0&(1-\alpha)\sigma_2\end{array}\right)\partial_x+ \widetilde{\mathbb{V}} \, , \quad \widetilde{\mathbb{V}}=\left(\begin{array}{cc}\widetilde{V}_{11}&\widetilde{V}_{12}\\\widetilde{V}_{12}^\dagger&\widetilde{V}_{22}\end{array}\right)+\frac{m\alpha}{2}\sigma_3\otimes\sigma_0,\nonumber\\
\label{H-alpha-R2}
\end{eqnarray}
where
\begin{eqnarray}
\widetilde{V}_{11}&=&(1+\alpha)\left(\frac{m }{2}\sigma_3+\left(-\frac{m}{2}\frac{ u_{22}}{u_{12}}+v_1\right)\sigma_1\right),\\
\widetilde{V}_{22}&=&\widetilde{V}_{11}\vert_{\alpha\rightarrow-\alpha},\\
\widetilde{V}_{12}&=&-\frac{m}{2}\left(\sigma_{3}+\frac{u_{22}}{u_{12}}(\sigma_{1}+i\alpha\sigma_{2})\right).
\end{eqnarray}
The two Dirac fermions differ both in their Fermi velocities as well as in their masses. 

\section{Solvability and spectral isomorphism}
\label{sec:Ext-Spec}

The equation (\ref{Halphastac}) suggests that the Hamiltonian $\widetilde{\mathbb{H}}_\alpha$ for $\alpha\neq0$ can have spectral properties quite different from those of $H$ and $\widetilde{H}$. We devote this section to a detailed analysis of its formal solutions. In particular, we discuss the form or general eigenstates corresponding to real eigenvalues, level crossing and creation of bound states in the continuum (BICs). 

\subsection{Fundamental sets of solutions for real eigenvalues of $\mathbb{H}_\alpha $}

\begin{figure}
	\centering
	
	\subfloat[][]{\includegraphics[width=0.35\textwidth]{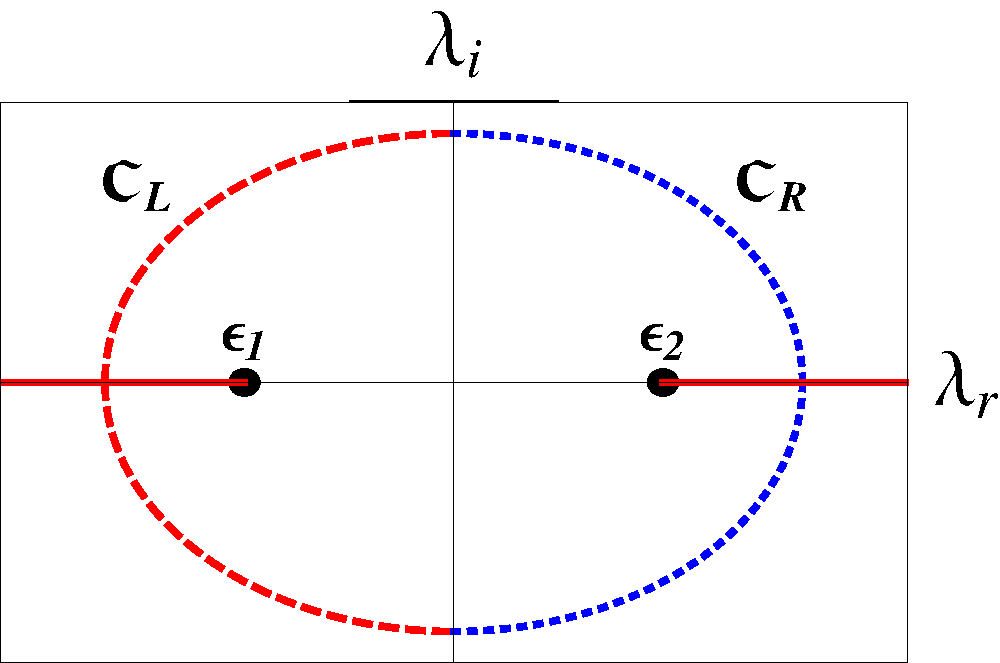}
	}
	\hspace{2mm}
	\subfloat[][]{\includegraphics[width=0.35\textwidth]{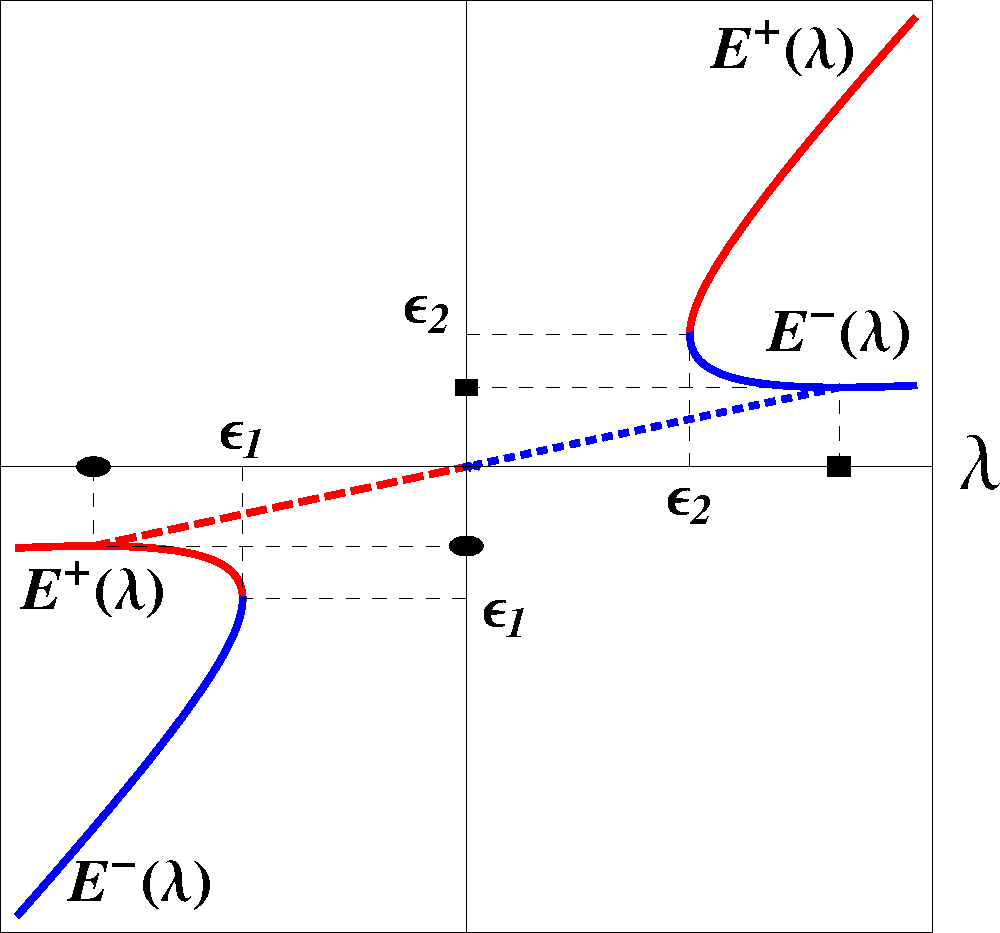}
}
	\caption{(a) Complex $\lambda$-contours $\mathfrak{C}_{L}$ (long-dashed-red) $\mathfrak{C}_{R}$ (short-dashed-blue), and $\mathfrak{L}$ (red), with $\lambda=\lambda_{r}+i\lambda_{i}$. (b) Real eigenvalues of the composite system $\mathbb{H}_{\alpha}$ evaluated in the appropriate contours, i.e., $\mathbb{E}^{+}(\lambda)\vert_{\mathfrak{L}}$ (thick red), $\mathbb{E}^{-}(\lambda)\vert_{\mathfrak{L}}$ (thick blue) $\mathbb{E}^{+}(\lambda)\vert_{\mathfrak{C}_{L}}$ (long-dashed-red), and $\mathbb{E}^{-}(\lambda)\vert_{\mathfrak{C}_{R}}$ (dotted blue). The small ellipses and rectangles denote the coordinates $\{\lambda_{\uparrow},\mathbb{E}_{\uparrow}\}$ $\{\lambda_{\downarrow},\mathbb{E}_{\downarrow}\}$ of the functions $\mathbb{E}^{+}(\lambda)$ and $\mathbb{E}^{-}(\lambda)$, respectively, as given in~\eqref{Ext-equilibrium}. In both cases, the parameters have been fixed as $\alpha=0.5$ and $\epsilon_{1}=-\epsilon_{2}=-2$.}
	\label{fig:Ext-spec}
\end{figure}

The equation 
\begin{equation}
{\mathbb{H}}_{\alpha}f(x)=\mathbb{E}^\pm(\lambda) f(x)
\end{equation}
is a system of four  coupled ordinary differential equations of the first order. Therefore, it has four independent solutions for any complex $\mathbb{E}^{\pm}(\lambda)$. Let us focus on the situations where $\mathbb{E}^{\pm}(\lambda)$ acquires real values. We have
	\begin{equation}\label{ee}
	\mathbb{E}^\pm(\lambda)=\lambda\pm\alpha\sqrt{(\lambda-\epsilon_{1})(\lambda-\epsilon_{2})}\in\mathbb{R}\quad \mbox{for}\quad  \lambda\in\mathfrak{L}=(-\infty,\epsilon_{1}]\cup[\epsilon_{2},\infty) \, .
	\end{equation} 
The curves $\mathbb{E}^\pm(\lambda)$ span conic sections for  $\lambda\in\mathfrak{L}$ such that they form together two hyperbolas, see Fig.~\ref{fig:Ext-spec}. These tend to to the straight lines asymptotically, \begin{equation}\mathbb{E}^{\pm}(\lambda)=\begin{cases}(1\pm\alpha)\lambda+o(\lambda),\quad \lambda\rightarrow\infty,\\ (1\mp\alpha)\lambda+o(\lambda),\quad \lambda\rightarrow-\infty.\end{cases}\label{asymptotics}\end{equation}  
The hyperbola corresponding to $\lambda\leq \epsilon_1$  acquires its maximum $\mathbb{E}_\uparrow$ at $\lambda_\uparrow$. The hyperbola plotted for $\lambda\geq\epsilon_2$ has its minimum $\mathbb{E}_\downarrow$ at $\lambda_{\downarrow}$. They are explicitly given by 
	\begin{equation}
	\begin{aligned}
	&\lambda_{\uparrow}=\Delta-\frac{\epsilon_{2}-\epsilon_{1}}{2\sqrt{1-\alpha^{2}}} \, , \quad 
	\mathbb{E}_{\uparrow}=\Delta-\frac{\sqrt{1-\alpha^{2}}}{2}(\epsilon_{2}-\epsilon_{1}) \, , \\
	&\lambda_{\downarrow}=\Delta+\frac{\epsilon_{2}-\epsilon_{1}}{2\sqrt{1-\alpha^{2}}} \, , \quad 
	\mathbb{E}_{\downarrow}=\Delta+\frac{\sqrt{1-\alpha^{2}}}{2}(\epsilon_{2}-\epsilon_{1}) \, ,\quad \Delta=\frac{\epsilon_{1}+\epsilon_{2}}{2}.
	\end{aligned}
	\label{Ext-equilibrium}
	\end{equation}
	The energy of the extended Hamiltonian is thus defined through the isomorphism (\ref{ee}), $\mathbb{E}^{\pm}(\lambda):\mathfrak{L}\mapsto (-\infty,\mathbb{E}_{\uparrow}]\cup[\mathbb{E}_{\downarrow},\infty)$.

It is impossible to get $\mathbb{E}^{\pm}(\lambda)\in(\mathbb{E}_{\uparrow},\mathbb{E}_{\downarrow})$ for real $\lambda$. We can extend $\lambda$ into a specific contour in the complex plane $\mathfrak{C}$ for which $\mathbb{E}^{\pm}(\lambda)\vert_{\lambda\in\mathfrak{C}}$ is still real. It is possible to show after some calculations that the condition Im$[\mathbb{E}^{\pm}(\lambda_{r}+i\lambda_{i})]=0 $, where $\lambda_r$ and $\lambda_i$ are real,  can be satisfied provided that $\lambda$ belongs to the following contour $\mathfrak{C}=\mathfrak{C}_{L}\cup\mathfrak{C}_{R}$, where
	\begin{equation}
	\begin{aligned}
	&\mathfrak{C}_{L}=\left\{ \lambda_{r}+i\lambda_{i} \mid \frac{4(1-\alpha^{2})}{(\epsilon_{1}-\epsilon_{2})^{2}}\left( \lambda_{r} - \frac{\epsilon_{1}+\epsilon_{2}}{2} \right)^{2}+\frac{4(1-\alpha^{2})}{\alpha^{2}(\epsilon_{1}-\epsilon_{2})^{2}}\lambda_{i}^{2}=1,
	\, \lambda_{r}\in\left(\lambda_{\uparrow},\Delta \right) \right\} \, , \\
	&\mathfrak{C}_{R}=\left\{ \lambda_{r}+i\lambda_{i} \mid \frac{4(1-\alpha^{2})}{(\epsilon_{1}-\epsilon_{2})^{2}}\left( \lambda_{r} - \frac{\epsilon_{1}+\epsilon_{2}}{2} \right)^{2}+\frac{4(1-\alpha^{2})}{\alpha^{2}(\epsilon_{1}-\epsilon_{2})^{2}}\lambda_{i}^{2}=1, \, \lambda_{r}\in\left( \Delta ,\lambda_{\downarrow}\right) \right\} \, .
	\end{aligned}
	\end{equation}
	The contours $\mathfrak{C}_{L}$ and $\mathfrak{C}_{R}$ span the left and right halves of an ellipse. From the latter, we obtain the mappings $\mathbb{E}^{+}(\lambda):{\mathfrak{C}_{L}}\mapsto(\mathbb{E}_{\uparrow},\Delta)$ and $\mathbb{E}^{-}(\lambda):{\mathfrak{C}_{L}}\mapsto(\Delta,\mathbb{E}_{\downarrow})$, i.e. we can recover any $\mathbb{E}^{\pm}(\lambda)\in(\mathbb{E}_{\uparrow},\mathbb{E}_{\downarrow})$ for $\lambda\in\mathfrak{C}_L\cup\mathfrak{C}_R$. In particular, there holds 
	\begin{equation}
	\mathbb{E}^{\pm}\left(i\,\lambda_i\right)=\Delta,\quad \lambda_i=1-\frac{(1-\alpha^2)(\epsilon_1+\epsilon_2)^2}{(\epsilon_1-\epsilon_2)^2}.
	\end{equation}
To illustrate our results, in Fig.~\ref{fig:Ext-spec} we depict the three contours $\mathfrak{C}_{L}$, $\mathfrak{C}_{R}$, and $\mathfrak{L}$ and we show the corresponding energies associated to each contour. 

Now, we can discuss the properties of the fundamental set of solutions of (\ref{Halphastac}) for any real eigenvalue. It is convenient to denote $\psi_a(\lambda)$, $a=1,2$, the two independent solutions of $(H-\lambda)\psi_a(\lambda)=0$. Correspondingly, we denote 
\begin{equation}\Psi_{a}^{\pm}(\lambda)=\left(\begin{array}{c}\pm\sqrt{F(\lambda)}\psi_a(\lambda)\\L\psi_a(\lambda)\end{array}\right),\quad a=1,2.\label{Psi_a}\end{equation}

We split our discussion into the following cases:

$\bullet$ For any $\mathbb{E}\in(\epsilon_2,\infty)$, there exist $\lambda_+,\lambda_-\in(\epsilon_2,\infty)$ such that $\mathbb{E}^{+}(\lambda_+)=\mathbb{E}^{-}(\lambda_-)=\mathbb{E}$. The four independent eigenstates of $\mathbb{H}_{\alpha}$ corresponding to this eigenvalue are
\begin{equation}
(\mathbb{H}_{\alpha}-\mathbb{E})\Psi_{a}^{\pm}(\lambda_{\pm})=0, \quad a=1,2.
\end{equation}

$\bullet$ For any $\mathbb{E}\in(\mathbb{E}_{\downarrow},\epsilon_2)$, there exist $\lambda,\kappa\in[\epsilon_2,\infty)$ such that $\mathbb{E}^{-}(\lambda)=\mathbb{E}^{-}(\kappa)=\mathbb{E}$, and the corresponding eigenvectors are
\begin{equation}
(\mathbb{H}_{\alpha}-\mathbb{E})\Psi_{a}^{-}(\lambda)=0,\quad (\mathbb{H}_{\alpha}-\mathbb{E})\Psi_{a}^{-}(\kappa)=0,\quad a=1,2.
\end{equation}  

$\bullet$ When $\mathbb{E}=\epsilon_2$, there is $\lambda\in(\epsilon_2,\infty)$ such that $\mathbb{E}^{-}(\lambda)=\epsilon_2$. The four independent vectors corresponding to this eigenvalue are
\begin{equation}
\Psi_a^{-}(\lambda),\quad\Psi_3=\left(\begin{array}{c}u_2\\0\end{array}\right),\quad \Psi_4=\left(\begin{array}{c}0\\\tilde{u}_2\end{array}\right),\quad a=1,2.
\end{equation}

$\bullet$ For any $\mathbb{E}\in(-\infty,\epsilon_1)$, there exist $\lambda_-,\lambda_+\in(-\infty,\epsilon_1)$ such that  $\mathbb{E}^{-}(\lambda_-)=\mathbb{E}^{+}(\lambda_+)=\mathbb{E}$, respectively. Likewise in the previous case, the four solutions are
\begin{equation}
(\mathbb{H}_{\alpha}-\mathbb{E})\Psi_{a}^{\pm}(\lambda_{\pm})=0,\quad \lambda_{\pm}\in(-\infty,\epsilon_1) \quad a=1,2.
\end{equation}

$\bullet$ When $\mathbb{E}\in(\epsilon_1,\mathbb{E}_{\uparrow})$, then there exist $\lambda,\kappa\in(-\infty,\epsilon_1)$ such that $\mathbb{E}^+(\lambda)=\mathbb{E}^+(\kappa)=\mathbb{E}$. The corresponding eigenstates are
\begin{equation}
(\mathbb{H}_{\alpha}-\mathbb{E})\Psi_{a}^{+}(\lambda)=0,\quad (\mathbb{H}_{\alpha}-\mathbb{E})\Psi_{a}^{+}(\kappa)=0,\quad a=1,2.
\end{equation} 

$\bullet$ When $\mathbb{E}=\epsilon_1$, there is $\lambda\in(-\infty,\epsilon_1)$ such that $\mathbb{E}^{+}(\lambda)=\epsilon_1$. The eigenvectors corresponding to this eigenvalue read as
\begin{equation}
\Psi_a^{+}(\lambda),\quad\Psi_3=\left(\begin{array}{c}u_1\\0\end{array}\right),\quad \Psi_4=\left(\begin{array}{c}0\\\tilde{u}_1\end{array}\right),\quad a=1,2.
\end{equation}

$\bullet$ Let us consider $\mathbb{E}=\{\mathbb{E}_{\uparrow},\mathbb{E}_{\downarrow}\}$. Recall that these eigenvalues correspond to the extremal points of the conic curves, i.e., $\frac{d\mathbb{E}^{\pm}(\lambda)}{d\lambda}\vert_{\lambda_{\downarrow}}=\frac{d\mathbb{E}^{\pm}(\lambda)}{d\lambda}\vert_{\lambda_{\uparrow}}=0$. There are two solutions $\Psi_{a}^{-}(\lambda_{\downarrow})$ and $\Psi_a^+(\lambda_\uparrow)$ for each of $\mathbb{E}_{\downarrow}$, $\mathbb{E}_{\uparrow}$, respectively,  defined as in (\ref{Psi_a}). In order to find the remaining two solutions for each eigenvalue, we exploit the eigenvalue equation $\mathbb{H}_{\alpha}\Psi^{\pm}_a(\lambda)=\mathbb{E}^{\pm}(\lambda)\Psi^{\pm}_a(\lambda)$, with $\Psi^{\pm}_a(\lambda)$ independent solutions for a fixed $\lambda$,  $a=1,2$. By differentiating the eigenvalue equation with respect to $\lambda$, and recalling that $\mathbb{H}_{\alpha}$ is independent of $\lambda$, we obtain $(\mathbb{H}_{\alpha}-\mathbb{E}^{\pm}(\lambda))\frac{d}{d\lambda}(\Psi^{\pm}_a(\lambda))=\frac{d\mathbb{E}^{\pm}(\lambda)}{d\lambda}\Psi^{\pm}_a(\lambda)$. Since $\frac{d\mathbb{E}^{\pm}(\lambda)}{d\lambda}=0$ at the points $\lambda_{\downarrow}$ and $\lambda_{\uparrow}$, we conclude that $\frac{d}{d\lambda}\Psi^{+}_a(\lambda)\vert_{\lambda_{\uparrow}}$ and $\frac{d}{d\lambda}\Psi^{-}_a(\lambda)\vert_{\lambda_{\downarrow}}$ are also eigensolutions of $\mathbb{H}_{\alpha}$ with eigenvalues $\mathbb{E}_{\uparrow}$ and $\mathbb{E}_{\downarrow}$, respectively. In this form, we found the remaining unknown solutions.

$\bullet$ For each $\mathbb{E}\in(\mathbb{E}_{\uparrow},\frac{\epsilon_1+\epsilon_2}{2}]$, there exist $\lambda,\lambda^*\in\mathfrak{C}_L$ such that $\mathbb{E}^{+}(\lambda)=\mathbb{E}^{+}(\lambda^*)=\mathbb{E}.$ The corresponding eigenstates are $\Psi_a^{+}(\lambda)$ and $\Psi_a^+(\lambda^*)$. For any $\mathbb{E}\in[\frac{\epsilon_1+\epsilon_2}{2},\mathbb{E}_{\downarrow})$, there are $\lambda,\lambda^*\in\mathfrak{C}_R$ such that $\mathbb{E}^{-}(\lambda)=\mathbb{E}^{-}(\lambda^*)=\mathbb{E}.$ The four eigenvectors are $\Psi_a^{-}(\lambda)$ and $\Psi_a^-(\lambda^*)$.

\subsection{Level crossing }
\label{sec:crossing}
Let us fix four eigenvalues of $\mathbb{H}_0$, $\lambda_1,\lambda_2\in(-\infty,\epsilon_1]$ and $\lambda_3,\lambda_4\in[\epsilon_2,\infty)$, such that $\lambda_1<\lambda_2<\lambda_3<\lambda_4$. For each $\lambda_a$, $a=1,\dots,4,$ there are two eigenvalues $\mathbb{E}^{\pm}(\lambda_a)$ of $\mathbb{H}_\alpha$. They have linear dependence on $\alpha$: $\mathbb{E}^+(\lambda_a)$ are growing functions of $\alpha$ whereas $\mathbb{E}^-(\lambda_a)$ are decreasing functions of $\alpha$. Therefore, there occurs level crossing for specific values of $\alpha$. Let us discuss this situation in more detail here. 

As we discussed in (\ref{Ext-equilibrium}), the hyperbola formed by $\mathbb{E}^{\pm}(\lambda)$ for $\lambda\leq\epsilon_1$ has the maximum $\mathbb{E}^+(\lambda_{\uparrow})=\Delta-\frac{\sqrt{1-\alpha^{2}}}{2}(\epsilon_{2}-\epsilon_{1}) $ whereas  the hyperbola formed by $\mathbb{E}^{\pm}(\lambda)$ for $\lambda\geq\epsilon_2$ has the minimum $\mathbb{E}^-(\lambda_{\downarrow})=\Delta+\frac{\sqrt{1-\alpha^{2}}}{2}(\epsilon_{2}-\epsilon_{1}) $. We can see that the two values tend to the same constant $\Delta$ for $\alpha\rightarrow1$, one from above while the other one from below. Therefore, $\mathbb{E}^{\pm}(\lambda_{1(2)})$ never cross with $\mathbb{E}^{\pm}(\lambda_{3(4)})$ for any $\alpha\in[0,1)$.

There are two possible types of level crossings of $\mathbb{E}^{\pm}(\lambda_{3(4)})$: 
\begin{align}
&\mathbb{E}^+(\lambda_3)=\mathbb{E}^{-}(\lambda_4)\Leftrightarrow\alpha^{+-}(\lambda_3,\lambda_4)=\frac{\lambda_4-\lambda_3}{\sqrt{F(\lambda_3)}+\sqrt{F(\lambda_4)}},\label{levelcrossing0}\\
&\mathbb{E}^-(\lambda_3)=\mathbb{E}^{-}(\lambda_4)\Leftrightarrow\alpha^{--}(\lambda_3,\lambda_4)=\frac{\lambda_4-\lambda_3}{\sqrt{F(\lambda_4)}-\sqrt{F(\lambda_3)}},\label{levelcrossing1}\\
&\lambda_3<\lambda_4,\quad \lambda_3,\lambda_4\in[\epsilon_2,\infty).\nonumber
\end{align}
These formulas can be obtained directly from (\ref{Halphastac}).  
There holds  $\alpha^{+-}(\lambda_3,\lambda_4)>0$ and $\alpha^{--}(\lambda_3,\lambda_4)>0$. It follows from the fact that $F(\lambda)$ is growing function of $\lambda>\Delta$. It is rather straightforward to see that $\alpha^{+-}(\lambda_3,\lambda_4)<1$. We also have $\alpha^{--}(\lambda_3,\lambda_4)<1$. Indeed, there holds $\mathbb{E}^{-}(\lambda_3)<\mathbb{E}^-(\lambda_4)$ for sufficiently small $\alpha>0$. Additionally, there exists $\alpha<1$ such that $\mathbb{E}^-(\lambda)$ acquires its minimum at $\lambda_4$, $\mathbb{E}^-(\lambda_4)=\mathbb{E}_{\downarrow}<\mathbb{E}^-(\lambda_3)$. Therefore, there must exist $\alpha^{--}(\lambda_3,\lambda_4)\in(0,1)$ such that (\ref{levelcrossing1}) is satisfied.    

The level crossing of $\mathbb{E}^{\pm}(\lambda_{1(2)})$ can be discussed in the same vain. We get
\begin{align}
&\mathbb{E}^+(\lambda_1)=\mathbb{E}^{-}(\lambda_2)\Leftrightarrow\alpha^{+-}(\lambda_1,\lambda_2)=\frac{\lambda_2-\lambda_1}{\sqrt{F(\lambda_1)}+\sqrt{F(\lambda_2)}},\\
&\mathbb{E}^-(\lambda_1)=\mathbb{E}^{-}(\lambda_2)\Leftrightarrow\alpha^{++}(\lambda_1,\lambda_2)=\frac{\lambda_2-\lambda_1}{\sqrt{F(\lambda_2)}-\sqrt{F(\lambda_1)}},\\
&\lambda_1<\lambda_2,\quad \lambda_1,\lambda_2\in(-\infty,\epsilon_1].\label{levelcrossing2}
\end{align} 
Likewise in the previous case, we have $\alpha^{+-}(\lambda_1,\lambda_2),\alpha^{++}(\lambda_1,\lambda_2)\in(0,1)$ that can be obtained with the use of the same arguments as in (\ref{levelcrossing1}).

Let us suppose that $\lambda_3$ and $\lambda_4$ belong to the energy spectrum of $\mathbb{H}_0$. We denote the corresponding physical states as $\Psi_{\lambda_3}$ and $\Psi_{\lambda_4}$. The eigenstates $\Psi_{\lambda_3}$ have lower energy than $\Psi_{\lambda_4}$ for $\alpha=0$. However, it follows from (\ref{levelcrossing1}) that there exists $\alpha\in(0,1)$ such that $\Psi_{\lambda_3}$ has higher energy $\Psi_{\lambda_4}$ in the system described by $\mathbb{H}_{\alpha}$. It demonstrates that the oscillation theorem known for Schr\"odinger operators does not hold for this system.

\subsection{Bound states in the continuum (BICs)}
Let us suppose that the spectrum of $\mathbb{H}_0$ consists of two bands of continuous energies divided by a gap that contains one discrete energy level $\epsilon$,  \begin{equation}\sigma(\mathbb{H}_0)=(-\infty,m_1]\cup[m_2,\infty)\cup\{\epsilon\},\quad \frac{\epsilon_1+\epsilon_2}{2}<\epsilon<m_2.\end{equation}
As $\alpha$ increases, the spectral gap of $\mathbb{H}_{\alpha}$ gets closed so that the discrete energies get immersed into the continuous spectrum for a critical value of $\alpha$ and the corresponding eigenstate becomes the bound state in the continuum. Let us consider this situation in more detail. 

The spectrum of $\mathbb{H}_{\alpha}$ is
\begin{equation}
\sigma(\mathbb{H}_\alpha)=(-\infty,\mathbb{E}_{max}]\cup[\mathbb{E}_{min},\infty)\cup\{\mathbb{E}^+(\epsilon),\mathbb{E}^-(\epsilon)\}.
\end{equation}
The thresholds $\mathbb{E}_{max}$ and $\mathbb{E}_{min}$  of the spectral bands are given as follows
\begin{align}
&\mathbb{E}_{max}=\max_{\lambda\in(-\infty,m_1]}\mathbb{E}^+(\lambda)=\begin{cases}\mathbb{E^+}(m_1),\quad m_1>\lambda_{\uparrow},\\
\mathbb{E}_{\uparrow},\quad m_1\leq \lambda_{\uparrow},  \end{cases}\\
&\mathbb{E}_{min}=\min_{\lambda\in[m_2,\infty)}\mathbb{E}^-(\lambda)=\begin{cases}\mathbb{E^-}(m_2),\quad m_2>\lambda_{\downarrow},\\
\mathbb{E}_{\downarrow},\quad m_2\leq \lambda_{\downarrow}.  \end{cases}\label{Emin}
\end{align}
The explicit expressions for $\mathbb{E}_{\uparrow\downarrow}$ and $\lambda_{\uparrow\downarrow}$ can be found in (\ref{Ext-equilibrium}). 

The bound state energies of  $\mathbb{E}^{\pm}(\epsilon)$ get immersed into the continuum when $\mathbb{E}^{\pm}(\epsilon)\geq\mathbb{E}_{max}$. We denote  $\alpha_{crit}^{\pm}(\epsilon)$ the critical value of $\alpha$ such that 
\begin{equation}
\mathbb{E}^{\pm}(\epsilon)=\mathbb{E}_{min}\vert_{\alpha=\alpha_{crit}^{\pm}(\epsilon)}.\label{alphacrit}
\end{equation}
Let us focus on $\alpha^+_{crit}(\epsilon)$. Taking into account the definition (\ref{Emin}), there can be two possible formulas for  $\alpha^+_{crit}(\epsilon)$.
In the first case, it should solve $\mathbb{E}^{+}(\epsilon)=\mathbb{E}_{\downarrow}$. However, this equation has negative solution  only. Therefore, $\alpha^+_{crit}(\epsilon)$ can be found as the solution of   $\mathbb{E}^{+}(\epsilon)=\mathbb{E}^-(m_2)$ that was discussed in (\ref{levelcrossing0}), 
\begin{equation}
\alpha_{crit}^{+}(\epsilon)=\alpha^{+-}(\epsilon,m_2).\label{BIC1}
\end{equation}
For $\alpha_{crit}^-(\epsilon)$, the equation $\mathbb{E}^{-}(\epsilon)=\mathbb{E}_{\downarrow}$ has the solution $\alpha=2\frac{\sqrt{F(\epsilon)}}{2\epsilon-\epsilon_1-\epsilon_2}$. Substituting it into $\lambda_{\downarrow}$, we get $\lambda_{\downarrow}=\epsilon<m$. Therefore, it does not lie in the interval specified by (\ref{Emin}). The solution of $\mathbb{E}^{-}(\epsilon)=\mathbb{E}^{-}(m)$ is
\begin{equation}
\alpha_{crit}^{-}(\epsilon)=\alpha^{--}(\epsilon,m),\label{BIC2}
\end{equation}
where $\alpha^{--}(\epsilon,m_2)$ is given in (\ref{levelcrossing1}). 
The embedding of the discrete energies into the continuous spectrum will be discussed in the next section on the explicit examples.

\section{Examples}\label{Examples}
Let us illustrate the construction and properties of the composite operator $\mathbb{H}_{\alpha}$ on explicit examples. We will consider two systems where the interaction is strongly localized and vanishes for large $|x|$. Therefore, there is continuum spectrum divided by a gap that contains some bound state energies. In the first case, $\widetilde{H}$ corresponds to Darboux-transformed free-particle Hamiltonian, which renders $\widetilde{\mathbb{H}_{\alpha}}$ reflectionless. In the second case, we deal with shape-invariant systems based on P\"oschl-Teller model. 

\subsection{Darboux-transformed free particle model}
We consider a model that preserves the chiral symmetry after the Darboux transformation has been performed. To this end, let us consider the free particle model 
\begin{equation}
H=-i\sigma_{2}\partial_{x}+m\sigma_{1} \, , \quad H\psi=\lambda\psi \, , \quad H\sigma_3\psi=-\lambda\sigma_3\psi \, .
\end{equation}
Clearly, this model does not support any bound states, for the general solutions are of the form
\begin{equation}
\psi=d_{1}e^{\kappa_{\lambda}x}
\begin{pmatrix}
1 \\ \frac{\kappa_{\lambda}+m}{\lambda}
\end{pmatrix}
+
d_{2}e^{-\kappa_{\lambda}x}
\begin{pmatrix}
1 \\ \frac{-\kappa_{\lambda}+m}{\lambda}
\end{pmatrix}
\, , \quad 
\kappa^{2}_{\lambda}=m^{2}-\lambda^{2}\, ,
\label{CHI-scatt-H0}
\end{equation}
with $\lambda<m$, and $d_{1}$ and $d_{2}$ arbitrary real constants. The spectrum of $H$ is formed by two semi-infinite bands divided by the spectral gap, $\sigma(H)=(-\infty,-m]\cup[m,\infty).$ We now exploit the framework discussed in (\ref{Hchiral1})-(\ref{tVchiral1}) and construct a transformation matrix $U=(u_1,\sigma_{3}u_1)$. The latter is invertible if det$(U_{1})\neq 0$ for $x\in\mathbb{R}$. One way to achieve this is by fixing $d_{1}=d_{2}=\frac{1}{2}$ such that
\begin{equation}
U=
\begin{pmatrix}
u_{11} & u_{11} \\ u_{12} & -u_{12}
\end{pmatrix}
\, , \quad u_{11}=\cosh(\kappa_{\epsilon_{1}}x) \, , \quad u_{2}=\frac{\kappa_{\epsilon_{1}}\sinh(\kappa_{\epsilon_{1}}x)+m\cosh(\kappa_{\epsilon_{1}}x)}{\epsilon_{1}} \, , 
\end{equation}
and det$(U)=-2u_{11}u_{12}$ is clearly nodeless. Such a transformation matrix leads to a regular intertwining operator $L$ and Hamiltonian $\widetilde{H}$, 
\begin{equation}
\widetilde{H}=-i\sigma_{2}\partial_{x}+\widetilde{V}(x)\sigma_{1} \, ,\quad L=\partial_x+\left(\begin{array}{cc}-\kappa_{\epsilon_1}\tanh ( \kappa_{\epsilon_1}x)&0\\0&-m+\frac{\epsilon_1^2}{m+\kappa_{\epsilon_1}\tanh(\kappa_{\epsilon_1}x)}\end{array}\right) 
\end{equation}
that admits two bound states at $\epsilon_{1}$ and $-\epsilon_{1}$. The potential and missing states are respectively given by
\begin{equation}
\widetilde{V}(x)=\frac{\epsilon_1}{m+\kappa_{\epsilon_1}\tanh \kappa_{\epsilon_1}x} +\kappa_{\epsilon_1}\tanh \kappa_{\epsilon_1}x\, , \quad 
\widetilde{u}_1=
\begin{pmatrix}
\operatorname{sech}(\kappa_{\epsilon_{1}}x) \\ \frac{\epsilon_{1}\operatorname{sech}(\kappa_{\epsilon_{1}}x)}{m+\kappa_{\epsilon_{1}}\tanh(\kappa_{\epsilon_{1}}x)}
\end{pmatrix} \, , \quad 
\widetilde{u}_2=\sigma_{3}\widetilde{u}_{1} \, .
\end{equation}
The new system represented by $\widetilde{H}$ inherits scattering properties of $H$, in particular, there is no backscattering caused by $\widetilde{V}$. It can be understood from the fact that scattering states of $\widetilde{H}$ are obtained from the plane waves of $H$ via the intertwining operator $L$. Nevertheless, the later operator cannot change momentum of the plane wave as the second term in $L$ tends to a constant matrix for large $|x|$. The reflectionless Dirac systems were discussed in \cite{Correa}. 

\begin{figure}
	\centering
	\includegraphics[width=0.4\textwidth]{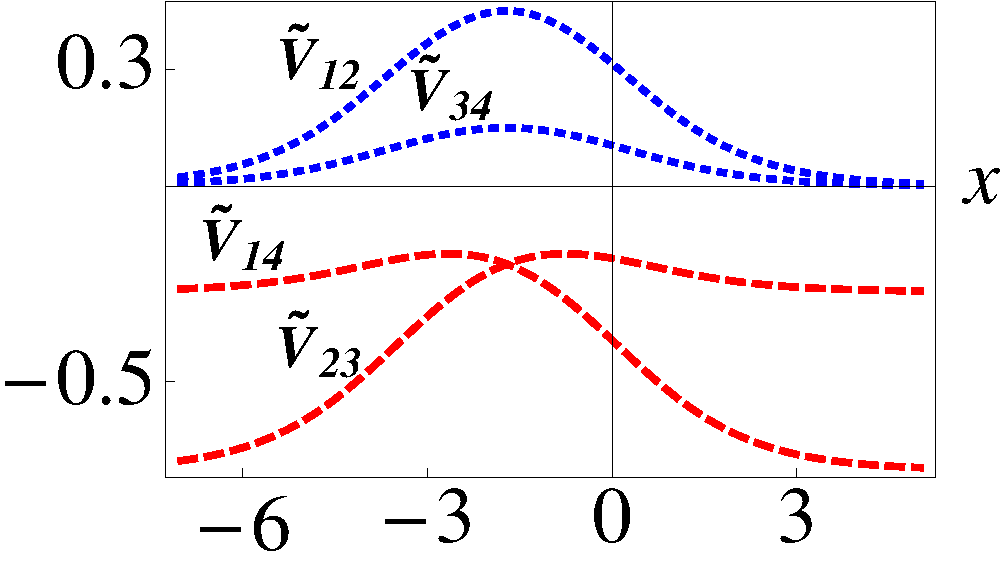}\hspace{3mm}	\includegraphics[width=0.4\textwidth]{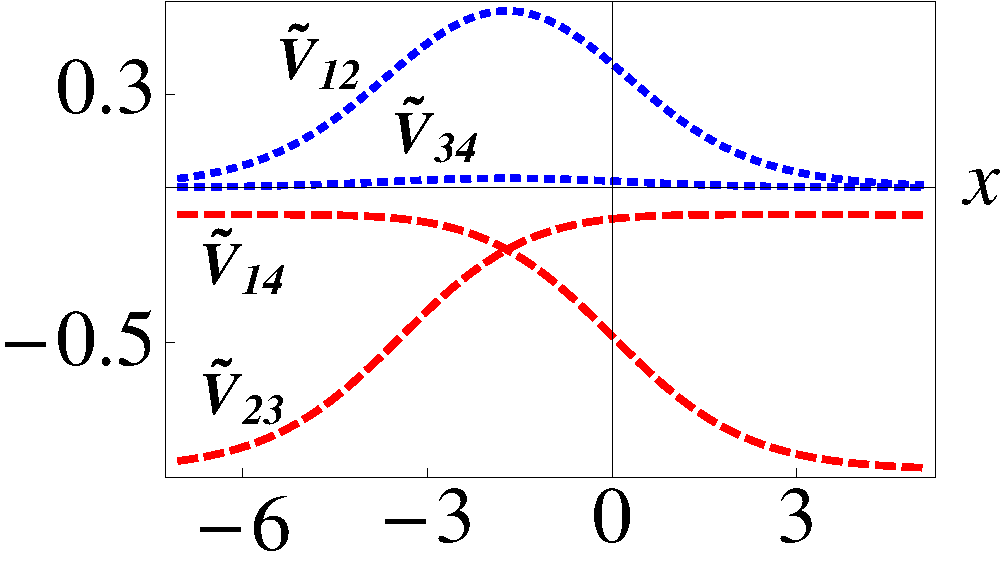}\hspace{3mm}\includegraphics[width=0.36\textwidth]{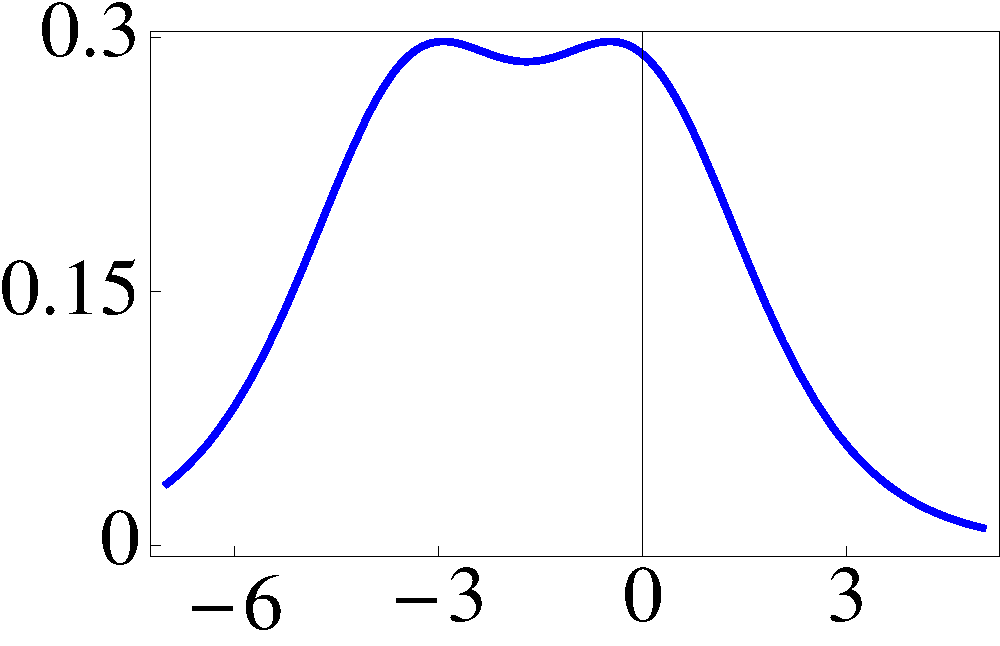}
	\caption{Components of the potential $\widetilde{\mathbb{V}}$ for $m=0.5$, $\epsilon_1=0.2$, $\alpha=0.5$ (left) and $\alpha=0.9$ (right). The bound states $\widetilde{u}_1$ and $\widetilde{u}_2$ do not depend on $\alpha$ and has the same density of probability. It is depicted on the lower figure.}
	\label{fig1}
	\end{figure}

Having derived the operator triplet $H$, $\widetilde{H}$ and $L$, we can construct the composite operator $\mathbb{H}_{\alpha}$. We illustrate here the explicit form of (\ref{tVchiral1}). The potential term 
\begin{equation}
\widetilde{\mathbb{V}}=U{\mathbb{V}}U^{-1}=\left(\begin{array}{cccc}0&\widetilde{\mathbb{V}}_{12}&0&\widetilde{\mathbb{V}}_{14}\\\widetilde{\mathbb{V}}_{12}&0&\widetilde{\mathbb{V}}_{23}&0\\0&\widetilde{\mathbb{V}}_{23}&0&\widetilde{\mathbb{V}}_{34}\\\widetilde{\mathbb{V}}_{14}&0&\widetilde{\mathbb{V}}_{34}&0\end{array}\right)
\end{equation}
has the following explicit realization,
\begin{align}
&\widetilde{\mathbb{V}}_{12}=\frac{(1+\alpha)\kappa^2_{\epsilon_1}}{m+m\cosh 2x\kappa_{\epsilon_1}+\kappa_{\epsilon_1}\sinh 2x\kappa_{\epsilon_1} },\quad \widetilde{\mathbb{V}}_{34}=\frac{1-\alpha}{1+\alpha }\widetilde{\mathbb{V}}_{12},\\
& \widetilde{\mathbb{V}}_{14}=-\mbox{sech}\kappa_{\epsilon_1}x\frac{\epsilon_1^2+(m^2(1+\alpha)-\alpha\epsilon_1^2)\cosh 2\kappa_{\epsilon_1}x+m(1+\alpha)\sinh 2\kappa_{\epsilon_1}x}{2(m\cosh \kappa_{\epsilon_1}x+\kappa_{\epsilon_1}\sinh\kappa_{\epsilon_1}x)},\\
&\widetilde{\mathbb{V}}_{23}=\widetilde{\mathbb{V}}_{14}\vert_{\alpha\rightarrow-\alpha}.
\end{align}
The components of the potential term tend to constant values in the limit of large $|x|$,
\begin{equation}
\lim_{x\rightarrow\pm\infty}\widetilde{\mathbb{V}}_{14}=-m-\frac{\alpha\kappa_{\epsilon_1}(\pm\kappa_{\epsilon_1}+m)}{\pm m+\kappa_{\epsilon_1}},\quad \lim_{x\rightarrow\pm\infty}\widetilde{\mathbb{V}}_{12}=0.
\end{equation}
The components of the potential term are plotted in Fig.~\ref{fig1} where is also the probability density associated with the missing state.

Let us analyze the spectrum of $\mathbb{H}_{\alpha}$ in dependence on $\alpha$.  For $\alpha=0$, there are two bands of energies divided by a gap that contains two discrete energies corresponding to the missing states,
\begin{equation}
\sigma(\mathbb{H}_{0})=(-\infty,-m]\cup[m,\infty)\cup\{-\epsilon_1,\epsilon_1\}.
\end{equation}
When we increase $\alpha$, the energy of the missing state remains the same, however, the threshold of the continuous spectrum gets altered. We get 
\begin{equation}
\sigma(\mathbb{H}_{\alpha})=(-\infty,-\mathbb{E}_{min}]\cup[\mathbb{E}_{min},\infty)\cup\{-\epsilon_1,\epsilon_1\}.
\end{equation}
The threshold value $\mathbb{E}_{min}$ is
\begin{equation}
\mathbb{E}_{min}=\begin{cases}
\mathbb{E}^-(m),\quad m>\lambda_{\downarrow},\\
\mathbb{E}_{\downarrow}=\epsilon_1\sqrt{1-\alpha^2},\quad m<\lambda_{\downarrow}.
\end{cases}\label{EminEmax}
\end{equation}
For increasing $\alpha$, the spectral gap gets closed and the energies of the missing states get immersed into the bands of the continuous spectrum. The missing states convert into the boundstates in the continuum (BICs) when $\alpha$ surpasses a critical value $\alpha_{crit}$. The critical value can be calculated from $\epsilon_1=\mathbb{E}_{min}\vert_{\alpha=\alpha_{crit}}$,
\begin{equation}
\alpha_{crit}=\sqrt{\frac{m-\epsilon_1}{m+\epsilon_1}}.
\end{equation}
In the Fig.\ref{alphacrit} , we illustrate the spectrum of $\mathbb{H}_{\alpha}$ for different values of $\alpha$. Revising the form of the general solutions of $(H-\lambda)\psi=0$ in (\ref{CHI-scatt-H0}), we conclude that there are no solutions with bounded density of probability for $\lambda\in\mathfrak{C}_L$ or $\lambda\in\mathfrak{C}_R$.

\begin{figure}
	\centering
\includegraphics[width=0.4\textwidth]{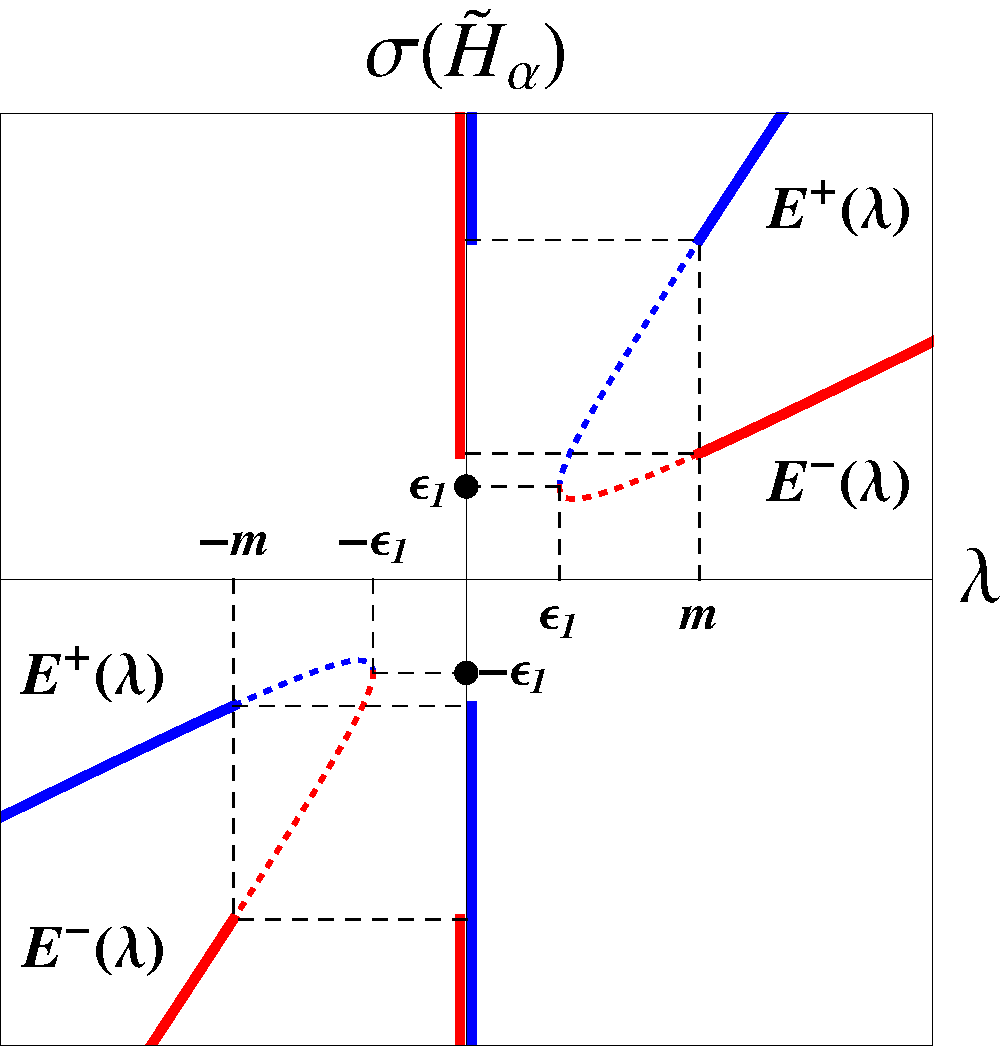}\hspace{8mm}	\includegraphics[width=0.4\textwidth]{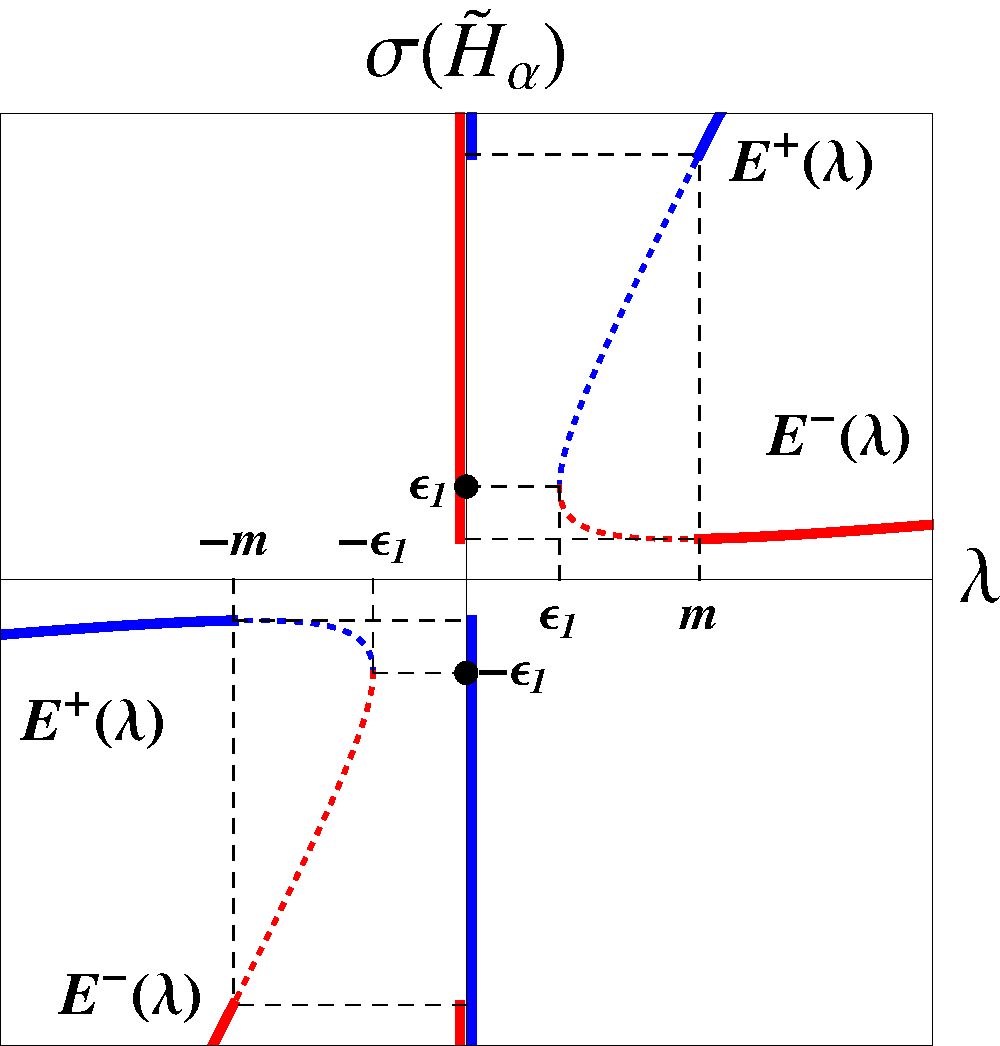}
		\caption{ The energy spectrum $\sigma(\mathbb{H}_{\alpha})$ (vertical axis) for  $m=0.5$, $\epsilon_1=0.2$, $\alpha=0.5$ (left) and $\alpha=0.9$ (right). The blue and red vertical lines correspond to the continuous spectrum (double degeneracy for each color, i.e. four-fold degeneracy in the overlap of red and black lines). The energies $\pm\epsilon_1$ of the bound states $\widetilde{u}_1$ and $\widetilde{u}_2=\sigma_3\widetilde{u}_1$ are represented by the black dot. They are discrete energies (left) or they correspond to BICs (right).  }
	\label{fig:EXT-FP}
\end{figure}

\begin{figure}
	\centering
	\includegraphics[width=0.4\textwidth]{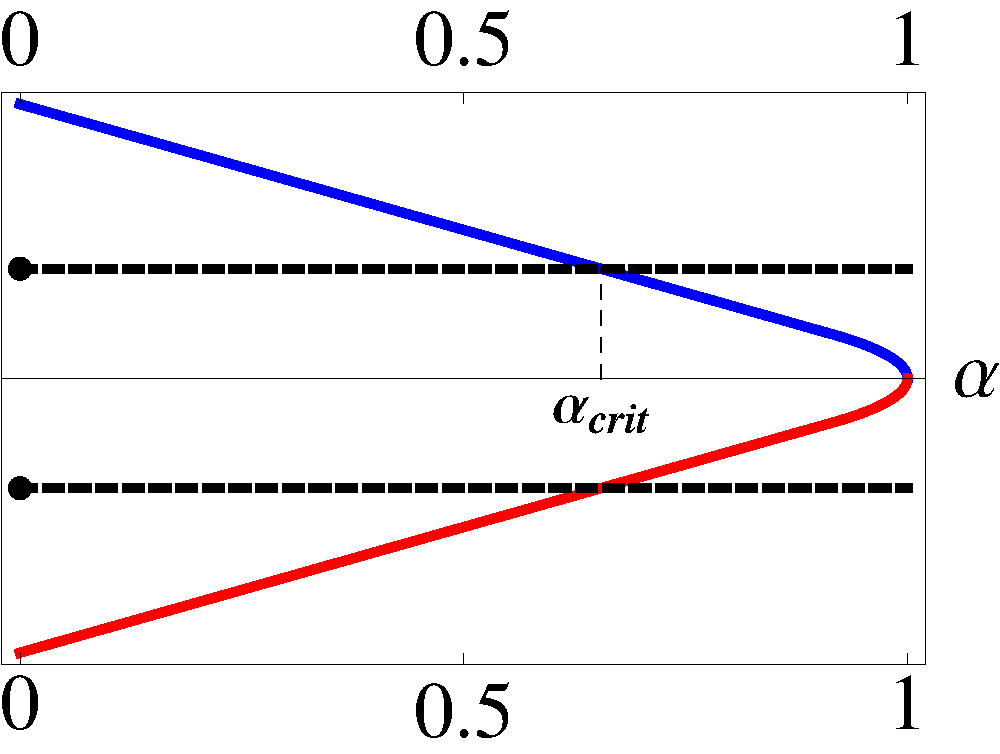}
	\label{fig:EXT-FP-c}
	\caption{Blue curve corresponds to $\mathbb{E}_{\min}$ whereas the red curve corresponds $\mathbb{E}_{max}=-\mathbb{E}_{\min}$, see (\ref{EminEmax}). The black dashed line corresponds to bound state energies $\pm\epsilon_1$ that get immersed into the continuous spectrum at $\alpha=\alpha_{crit}$. }
	\label{alphacrit}
\end{figure}

\subsection{P\"oschl-Teller-like interaction}
\label{sec:PT}
Alternatively, we may consider the second construction illustrated in Sec.~\ref{sec:susy-chiral}. To this end, we introduce an initial Hamiltonian that lacks chiral symmetry, defined by
\begin{equation}
	H=-i\sigma_{2}\partial_{x}+A_{\kappa}(x)\sigma_{1}+m\sigma_{3} \, , \quad A_{\kappa}(x)=U_{0}(\kappa-1)\tanh(U_{0}x) \, , \quad \kappa>1,\, U_{0}>0 \, ,
\end{equation}
where the mass term $m\sigma_{3}$ is responsible for the chiral symmetry breaking. In this case, the vector potential is asymptotically uniform, with a localized perturbation around the origin. 

The corresponding bound state eigensolutions of $H$ can be determined with ease by squaring $H$, leading to a diagonal operator that can be associated with the Schr\"odinger problem with a P\"oschl-teller interaction, $H^{2}=(-\partial_{xx}+A^{2}(x)+m^{2})\sigma_{0}-A'(x)\sigma_{3}$. The eigensolutions of the latter are well-known and can be obtained by taking the corresponding Sturm-Liovile problem into the \textit{hypergeometric form}~\cite{Nik88}. After some calculations one gets the bound state eigensolutions and eigenvalues
\begin{equation}
\begin{aligned}
& \psi_{n_{\pm}}^{(\pm)}=\mathcal{N}_{n_{\pm}}^{\pm}(1-y^{2}(x))^{\frac{\kappa-n_{\pm}-1}{2}}
\begin{pmatrix}
\mathcal{P}_{n_{\pm}}(x) \\
U_{0}\frac{2n_{\pm}y(x)\mathcal{P}_{n_{\pm}}(x)+(2\kappa-n_{\pm}-1)(1-y^{2}(x))\mathcal{P}_{n_{\pm}-1}(x)}{2(\lambda_{n}^{(\pm)}+m)}
\end{pmatrix}
\, , \\
& \lambda_{n_{\pm}}^{(\pm)}=\pm\sqrt{m^{2}-U_{0}^{2}n_{\pm}(n_{\pm}+2-2\kappa)} \, ,
\end{aligned}
\label{PT-eigensol}
\end{equation}
respectively, with $\mathcal{N}_{n}^{\pm}$ normalization factors, $y(x)=\tanh(U_{0}x)$, $\mathcal{P}_{n}(x)=P_{n}^{(a_{n},a_{n})}(y(x))$ the Jacobi polynomials~\cite{Olv10}, and $a_{n}=n-\kappa-1$. The indexes $n_{\pm}$ run over the values
\begin{equation}
\begin{cases}
n_{+}=0,\ldots, \lfloor \kappa-1\rfloor \, , \quad n_{-}=1,\ldots,\lfloor \kappa-1 \rfloor \quad & \kappa\in\mathbb{R}/\mathbb{Z}^{+} \\
n_{+}=0,\ldots, \kappa-2 \, , \quad n_{-}=1,\ldots, \kappa-2 & \kappa\in\mathbb{Z}^{+}
\end{cases}
\, .
\label{PT-index}
\end{equation}
As much as it can be obvious from the formula for $\lambda_{n_{\pm}}^{(\pm)}$, we stress that the indice $n_+$ counts the bound states with positive energy whereas $n_{-}$ denotes the bound states with negative energy. The lack of chiral symmetry is reflected by different range of the two indices.

Note that the term $(1-y^{2}(x))^{\frac{\kappa-n_{\pm}-1}{2}}=\operatorname{sech}(U_{0}x)^{\kappa-n-1}$ vanishes for $n_{\pm}=\lfloor \kappa-1 \rfloor$ when $\kappa\in\mathbb{Z}^{+}$, leading to non-vanishing eigensolutions $\psi_{n_{\pm}}^{(\pm)}$ for $\vert x\vert\rightarrow\infty$. Thus, such a case shall be excluded from the indexes, as done in~\eqref{PT-index}. On the other hand, $\psi_{0}^{-}$ given in~\eqref{PT-eigensol} is not well-defined for $n_{-}=0$. Despite the latter, the corresponding eigensolution $\psi_{0}^{(-)}$ can be computed directly from the stationary equation for $H$. This leads to $\psi_{0}^{(-)}\propto (\cosh(U_{0}x))^{\kappa-1}$, which clearly diverges asymptotically. For this reason, $n_{-}=0$ was excluded in~\eqref{PT-index}. Such a result reveals a lack of symmetry in the bound state eigenvalues, which is expected as $H$ is not chiral symmetric. 

The asymmetry in the physical and non-physical eigenvalues $\lambda_{0}^{+}=m$ and $\lambda_{0}^{(-)}=-m$, respectively, give us insight on how to select the seed solutions for Darboux transformation in order to get chiral-symmetric $\widetilde{H}$. To do so, it is immediate that one of the factorization energies should be $\epsilon_{1}=-m$, whereas the second factorization energy level shall be placed simultaneously at $\epsilon_{2}=0$. In this form, we bring symmetry into the energy eigenvalues and recast the desired chiral symmetry, which corresponds to the second scenario discussed in Sec.~\ref{sec:susy-chiral}. 

The transformation matrix $U$ has been determined in~\eqref{U2}, which, in the model under consideration, leads to the component $u_{21}=(\cosh(U_{0}x))^{\kappa-1}$. Although we can determine the remaining components in the most general case, we prefer to focus our discussion to a specific system where the formulas acquire remarkably simple form. To do so, we fix the mass term as $m=U_{0}\sqrt{2\kappa-1}$. The straightforward calculations lead to the transformation matrix
\begin{equation}
	U=
	\begin{pmatrix}
		0 & (\cosh(U_{0}x))^{\kappa} \\
		\cosh(U_{0}x)^{\kappa-1} & \sqrt{2\kappa-1}(\cosh(U_{0}x))^{\kappa}\tanh(U_{0}x)
	\end{pmatrix}
	\, , \quad 
	\Lambda=
	\begin{pmatrix}
		-U_{0}\sqrt{2\kappa-1} & 0 \\
		0 & 0
	\end{pmatrix}
	\, ,
\label{PT-U}
\end{equation}
such that det$(U)=-(\cosh(U_{0}x))^{2\kappa-1}\neq 0$ for $x\in\mathbb{R}$, as required. 

Therefore, the new Hamiltonian becomes
\begin{equation}
	\widetilde{H}=-i\sigma_{2}\partial_{x}+A_{\kappa+1}(x)\sigma_{1} \, ,
\end{equation}
which clearly has the desired chiral symmetry, $\{\widetilde{H},\sigma_{3}\}=0$.  It is remarkable that the $\widetilde{H}$ with the vector potential $A_{\kappa+1}(x)$ is the shape-invariant\footnote{Shape-invariant systems change only coupling constants when Darboux-transformed, see \cite{CooperKhare} for more details.} counterpart of $H$, yet, without the mass term. The bound state eigenfunctions of this Hamiltonian are computed from the kernel of the intertwining operator~\eqref{L2}, and from its action on~\eqref{PT-eigensol}. See~\ref{sec:DT}. Such an intertwining operator takes the form
\begin{equation}
L=\frac{d}{dx}+
\begin{pmatrix}
-A_{\kappa+1}(x) & 0 \\-U_{0}\sqrt{2\kappa-1} & -A_{\kappa}(x)
\end{pmatrix}
\label{PT-L}
\end{equation}
from which we construct the eigensolutions of $\widetilde{H}$. After some calculations we get
\begin{equation}
\begin{cases}
\widetilde{\psi}_{n}^{(\pm)}(x)=\mathcal{N}^{(\pm)}_{n}(1-y^{2}(x))^{\frac{\kappa-n-1}{2}}
\begin{pmatrix}
(n-2\kappa-1)\left( \frac{-2y(x)\mathcal{P}_{n}(x)+(1-y^{2}(x))\mathcal{P}_{n-1}(x)}{2} \right) \\
\mp \sqrt{(n+1)(n-2\kappa-1)}\mathcal{P}_{n}(x)
\end{pmatrix}  \\
\widetilde{\psi}_{\o}(x)=\sqrt{\frac{U_{0}\Gamma(\kappa+1/2)}{\Gamma(\kappa)\Gamma(1/2)}}(\operatorname{sech}(U_{0}x))^{\kappa}
\begin{pmatrix}
1 \\ 
0
\end{pmatrix} 
\end{cases}\label{widetildepsi}
\end{equation}
together with the corresponding energies
\begin{equation}
	\widetilde{\lambda}_{n}^{(\pm)}=\pm U_{0}\sqrt{(n+1)(2\kappa-n-1)} \, , \quad \widetilde{\lambda}_{\o}=0 \, , \quad n=0,\ldots,\widetilde{n}_{max} \, ,
\end{equation}
where 
\begin{equation}
	\begin{cases}
		\widetilde{n}_{max}=\lfloor \kappa-1\rfloor, \quad & \kappa\in\mathbb{R}/\mathbb{Z}^{+} \\
		\widetilde{n}=\kappa-2, \quad & \kappa\in\mathbb{Z}^{+}
	\end{cases}
	\, .
\end{equation}
In the latter, we have used $\widetilde{\psi}_{n}^{(\pm)}(x)=L\psi_{n}^{(\pm)}$ with $\psi_{n}^{(\pm)}(x)$ given in~\eqref{PT-eigensol}. It is worth to remark that both $\widetilde{\psi}_{0}^{(-)}(x)$ and $\widetilde{\psi}_{\o}(x)$ are the missing states added through the Daboux transform, which are not mapped by $L$. They thus play a special role in the composite construction. See details below.

Besides the set of bound-state eigensolutions previously determined, we may also explore the solutions associated with arbitrary eigenvalues. Of particular interest are the eigensolutions that behave asymptotically as plane waves, for they define scattering states. The corresponding eigenvalues are defined through $\lambda_{s}=\pm U_{0}\sqrt{\kappa^{2}+\nu_{s}^{2}}$, with $\nu_{s}\in\mathbb{R}$. Two fundamental solutions $\psi_{\nu_{s}}^{(j)}=(\psi_{1,\nu_{s}}^{(j)},\psi_{2,\nu_{s}}^{(j)})$, with $j=1,2$, can be identified for such scattering energies, the component of which are 
\begin{equation}
\begin{aligned}
&\psi_{1,\nu_{s}}^{(1)}=2^{-i\nu}(e^{U_{0}x}+e^{-U_{0}})^{i\nu_{s}} {}_{2}F_{1}\left( \left. \begin{aligned} 1-\kappa-i\nu_{s},\kappa-i\nu_{s} \\ 1-i\nu_{s} \hspace{10mm} \end{aligned} \right\vert \frac{1-\tanh(U_{0}x)}{2} \right) \, , \quad  \\
&\psi_{1,\nu_{s}}^{(2)}=2^{-i\nu_{s}}e^{-iU_{0}\nu_{s}x}{}_{2}F_{1}\left( \left. \begin{aligned} 1-\kappa,\kappa \\ 1+i\varepsilon \hspace{2mm} \end{aligned} \right\vert \frac{1-\tanh(U_{0}x)}{2} \right) \, ,\\
& \psi_{2,\nu_{s}}^{(j)}(x)=(\lambda_{s}+U_{0}\sqrt{2\kappa-1})^{-1}\left(\frac{d}{dx}\psi_{1,\nu_{s}}^{(j)}(x)+A_{\kappa}(x)\psi_{1,\nu_{s}}^{(j)}(x)\right).
\end{aligned}
\label{PT-H-scat1}
\end{equation}

Notice that the argument of the hypergeometric function in~\eqref{PT-H-scat1} becomes zero for $x\rightarrow \infty$, which reduces the hypergeometric function to one. In this limit, the positive exponential in $(e^{U_{0}x}+e^{-U_{0}x})$ dominates over the negative one. Then, both fundamental solutions become plane waves. For $x\rightarrow-\infty$, the argument in the hypergeometric function approaches to $1^{-}$, which is a well-known limit~\cite{Olv10} in which the hypergeometric function reduces to a complex constant\footnote{In this limit, the hypergeometric function reduces to a product Gamma functions, the specific form of which is irrelevant for the ongoing discussion. Further details can be found in~\cite{Olv10}.}. The solutions then reduce to plane waves as well. We can summary the asymptotic behavior as 
\begin{equation}
\begin{aligned}
&x\rightarrow+\infty: \quad \psi_{1;\nu_{s}}^{(1)}(x) \sim  e^{iU_{0}\nu_{s}x} \, , \quad \psi_{1;\nu_{s}}^{(1)}(x) \sim e^{iU_{0}\nu_{s}x} \, , \\
&x\rightarrow-\infty: \quad \psi_{1;\nu_{s}}^{(1)}(x) \sim  e^{-iU_{0}\nu_{s}x} \, , \quad \psi_{1;\nu_{s}}^{(1)}(x) \sim e^{-iU_{0}\nu_{s}x}.
\end{aligned}
\end{equation}
The same plane wave behavior is obtained for the components $\psi^{(j)}_{2,\nu_{s}}(x)$, for $j=1,2$. 

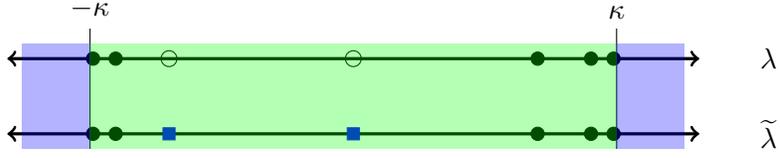
\begin{figure}
	\centering
	\begin{tikzpicture}
		\draw[<->,very thick] (-4.6,0)--(4.6,0) node[pos=1.1] {$\lambda$};
		\draw[-] (3.5,-1.2)--(3.5,0.4) node[above] {$\kappa$};
		\draw[-] (-3.5,-1.2)--(-3.5,0.4) node[above] {$-\kappa$};
		\filldraw[black] (-3.16,0) circle (2.5pt);
		\filldraw[black] (-3.46,0) circle (2.5pt);
		\draw (-2.45,0) circle (3pt);
		\draw (0,0) circle (3pt);
		\filldraw[black] (2.45,0) circle (2.5pt);
		\filldraw[black] (3.16,0) circle (2.5pt);
		\filldraw[black] (3.46,0) circle (2.5pt);
		\draw[<->,very thick] (-4.6,-1)--(4.6,-1) node[pos=1.1] {$\widetilde{\lambda}$};
		\filldraw[black] (-3.16,-1) circle (2.5pt);
		\filldraw[black] (-3.46,-1) circle (2.5pt);
		\filldraw[blue] (-2.53,-1.08) rectangle (-2.37,-0.92);
		\filldraw[blue] (-0.08,-1.08) rectangle (0.08,-0.92);
		\filldraw[black] (2.45,-1) circle (2.5pt);
		\filldraw[black] (3.16,-1) circle (2.5pt);
		\filldraw[black] (3.46,-1) circle (2.5pt);
		\path [fill=blue, opacity=0.3] (3.5,0.2) rectangle (4.4,-1.2);
		\path [fill=blue, opacity=0.3] (-3.5,0.2) rectangle (-4.4,-1.2);
		\path [fill=green, opacity=0.3] (3.5,0.2) rectangle (-3.5,-1.2);
	\end{tikzpicture}
	\caption{Point spectrum $\sigma(H)$ and $\sigma(\widetilde{H}$) related with the P\"oschl-Teller-like interaction and its Darboux transform, respectively, for $U_{0}=1$ and $\kappa=3.5$. Black dots denote the finite-norm solution eigenvalues, whereas hollow-dots denote the factorization eigenvalues $\epsilon_{1}$ and $\epsilon_{2}$ used in the Darboux transform. The blue and green areas denote the continuum and non-finite norm eigenvalue intervals, respectively, while the lines at $\lambda_{th}=\pm\kappa$ mark the eigenvalue thresholds.}
	\label{fig:PT-spec}
\end{figure}

One proceeds similarly with the Darboux transform $\widetilde{H}$, where the behavior of the solutions here is inherited from that of $H$. That is, the eigensolutions
\begin{equation}
\widetilde{\psi}^{(j)}_{\nu_{s}}=(\widetilde{\psi}_{1,\nu_{s}}^{(j)}, \widetilde{\psi}_{2,\nu_{s}}^{(j)})^{T}=L\psi^{(j)}_{\nu_{s}} \, ,
\end{equation}
behave asymptotically as plane waves as well. Thus, scattering solutions for $H$ and $\widetilde{H}$ exist simultaneously in the eigenvalue interval $\vert \lambda\vert>\vert\lambda_{th}\vert$, with $\lambda_{th}=\pm U_{0}\kappa$. This region is illustrated in Fig.~\ref{fig:PT-spec} as a blue-shadowed area.

In a similar fashion, there are eigenvalues $\lambda=\pm U_{0}\sqrt{k^{2}-\nu^{2}}$, with $-\kappa<\nu<\kappa$, such that the eigensolutions diverge asymptotically. These solutions lack any physical interpretation, contrary to the scattering or bound states. Still, as discussed previously, non-physical solutions are useful to construct well behaved (singularity-free) systems via Darboux transformation, provided that the required conditions are met. In this case, the solutions associated with the eigenvalues $\lambda=\epsilon_{1}=0$ and $\lambda=\epsilon_{2}=-m=-U_{0}\sqrt{2\kappa-1}$ have been considered. Moreover, note that $\lambda_{th}=\pm U_{0}\kappa$ defines the eigenvalue threshold that separates scattering solutions from the non-physical ones. The spectra of $H$ and $\widetilde{H}$ are depicted in Fig.~\ref{fig:PT-spec}, where the green area denotes the non-physical eigenvalues, the filled-circle the bound-state eigenvalues $\lambda_{n}$ and $\widetilde{\lambda}_{n}$, and the hollow-circle and filled-square the factorization eigenvalues $\epsilon_{1}$ and $\epsilon_{2}$.

\subsubsection*{Composite Hamiltonian $\mathbb{H}_{\alpha}$}
\begin{figure}
	\centering
	\subfloat[][$\alpha=0.25$]{\includegraphics[width=0.35\textwidth]{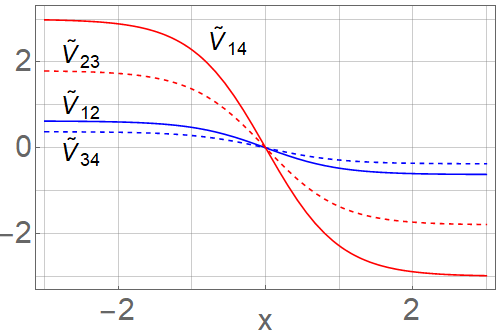}}
	\hspace{2mm}
	\subfloat[][$\alpha=0.9$]{\includegraphics[width=0.35\textwidth]{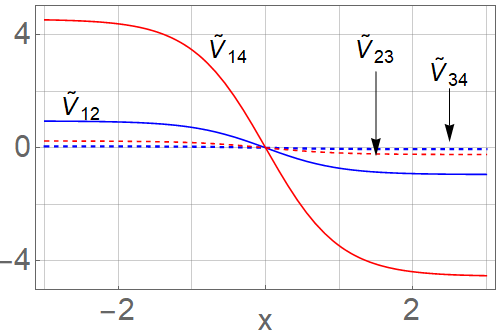}}
	\caption{Components of the matrix potential $\widetilde{\mathbb{V}}$ obtained ~\eqref{PT-V-alpha} for $\kappa=2.9$, $U_{0}=1$, and the indicated values of $\alpha$.}
	\label{fig:PT-compo-V}
\end{figure}

The particular choice of the transformation matrix $U$ in~\eqref{PT-U} allows us to obtain a relative simple expression for the $4\times 4$ matrix potential $\widetilde{\mathbb{V}}$ in~\eqref{H-alpha-R2}. We explicitly have
\begin{equation}
\mathbb{V}=\left(
\begin{array}{cccc}
\frac{\sqrt{2 k-1}(2 \alpha +1)}{2}  & -\frac{(\alpha +1)\tanh (x)}{2}  & -\frac{\sqrt{2 k-1}}{2} & -\frac{(2 k-1)(\alpha +1)\tanh (x)}{2} \\
 -\frac{(\alpha +1)\tanh (x)}{2}  & -\frac{\sqrt{2 k-1}}{2} & \frac{(2 k-1)(\alpha -1)\tanh (x)}{2}  & \frac{\sqrt{2 k-1}}{2} \\
 -\frac{\sqrt{2 k-1}}{2} & \frac{(2 k-1) (\alpha -1)\tanh (x)}{2} & \frac{\sqrt{2 k-1} (1-2 \alpha )}{2} & \frac{(\alpha -1)\tanh (x)}{2} \\
 -\frac{(2 k-1) (\alpha +1)\tanh (x)}{2}  & \frac{\sqrt{2 k-1}}{2} & \frac{(\alpha -1)\tanh (x)}{2}  & -\frac{\sqrt{2 k-1}}{2} \\
\end{array}
\right)
\, ,
\label{PT-V-alpha}
\end{equation}
For the case under consideration, we depict in Fig.~\eqref{fig:PT-compo-V} the non-uniform components of the matrix potential $\widetilde{\mathbb{V}}$ for two different values of $\alpha$ to illustrate the change on the interaction. Such components are asymptotically constant with smooth variations around the origin.

To construct the set of finite-norm solutions, let us first recall that the $\widetilde{\psi}_{0}^{(-)}(x)$ and $\widetilde{\psi}_{\o}(x)$ are the missing states associated with the energies $\epsilon_{1}=-U_{0}\sqrt{2\kappa-1}$ and $\epsilon_{2}=0$, respectively. Such energies play a special role in the spectrum of the composite Hamiltonian as they are invariant under changes of $\alpha$. The set of finite-norm solutions and their respective eigenvalues are thus summarized as
\begin{equation}
\begin{cases}
\Psi_{n;+}^{\pm}(x)=
\begin{pmatrix}
\pm\sqrt{F(\lambda_{n}^{(+)})} \, \psi_{n}^{(+)}(x) \\
L \psi_{n}^{(+)}(x)
\end{pmatrix}   \quad &\mathbb{E}^{\pm}(\widetilde{\lambda}_{n}^{(+)}) \, , \quad  n=0,\ldots, \widetilde{n}_{max} \\
\Psi_{n;-}^{\pm}(x)=
\begin{pmatrix}
\pm\sqrt{F(\widetilde{\lambda}^{(-)}_{n})} \, \psi_{n}^{(-)}(x) \\
L\psi_{n}^{(-)}(x)
\end{pmatrix}  \quad &\mathbb{E}^{\pm}(\widetilde{\lambda}_{n}^{(-)}) \, , \quad  n=1,\ldots, \widetilde{n}_{max} \\
\Psi_{0;-}(x)=
\begin{pmatrix}
0 \\
\widetilde{\psi}_{0}^{(-)}(x)
\end{pmatrix}  \quad &\mathbb{E}^{-}(\widetilde{\lambda}_{0}^{(-)})=\mathbb{E}^{+}(\widetilde{\lambda}_{0}^{(-)})=\widetilde{\lambda}_{0}^{(-)}=-U_{0}\sqrt{2\kappa-1} \\
\Psi_{\o}(x)=
\begin{pmatrix}
0 \\
\widetilde{\psi}_{\o}(x)
\end{pmatrix} \quad &\mathbb{E}^{-}(\widetilde{\lambda}_{\o})=\mathbb{E}^{+}(\widetilde{\lambda}_{\o})=\widetilde{\lambda}_{\o}=0 \\
\end{cases}
\end{equation}
where the eigenvalues $\mathbb{E}^{\pm}(\lambda)=\lambda\pm\alpha\sqrt{F(\lambda)}$, with $F(\lambda)=\lambda(\lambda+U_{0}\sqrt{2\kappa-1})$, were computed from~\eqref{Halphastac} and $\psi_{n}^{(\pm)}$ and $\widetilde{\psi}_{n}^{(\pm)}$ are given in (\ref{PT-eigensol}) and (\ref{widetildepsi}), respectively. 

\begin{figure}
	\centering
	\subfloat[][]{\includegraphics[width=0.3\textwidth]{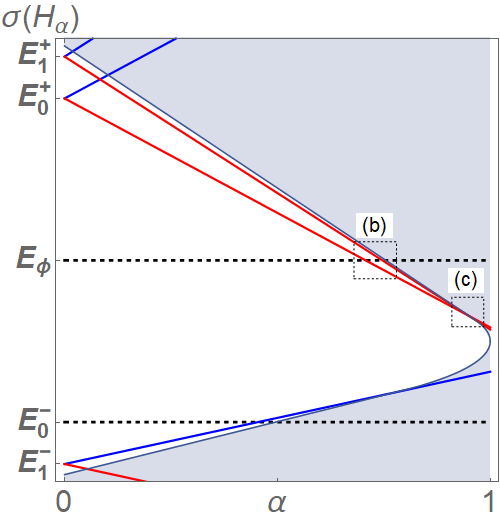}}
	\hspace{2mm}
	\subfloat[][]{\includegraphics[width=0.25\textwidth]{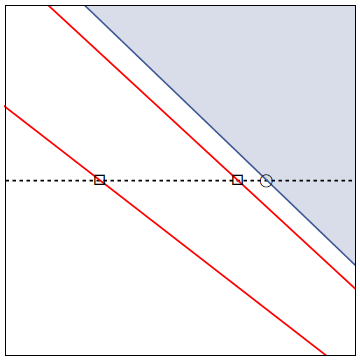}}
	\subfloat[][]{\includegraphics[width=0.25\textwidth]{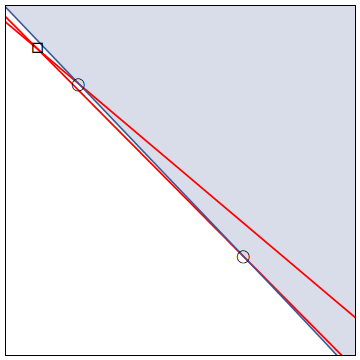}}
	\caption{(a) Eigenvalue curves $\mathbb{E}_{n}^{(+)}$ (blue) and $\mathbb{E}_{n}^{(-)}$ as functions of $\alpha\in(0,1)$. The shadowed area denotes the scattering regime. (b,c) Zoomed-in regions as marked in the dashed rectangles in (a). Square and circle markers indicate level crossings and transition into the continuum, respectively. In all cases, we have fixed $\kappa=2.9$ and $U_{0}=1$.}
	\label{fig:PT-compo-1}
\end{figure}

The spectral behavior of $\widetilde{\mathbb{H}}_{\alpha}$ is depicted in Fig.~\ref{fig:PT-compo-1} for $\kappa=2.9$ and $\lambda$ belonging to the finite-norm energies shown above, such that we have eight bound state eigenvalues. In addition, we have shadowed the regions in which the continuum spectrum exists. This provides a visual illustration about the transition of the bound states from the non-physical region into the continuum for the corresponding critical values of $\alpha$, which in turn reveals the appearance of BIC. Note that all the bound states get eventually trapped into the continuum, although the level crossings occur out of the continuum. This means that none of the BICs are degenerate, and only the bound states might become degenerate.  The critical values of $\alpha$ where either the level crossing occurs or the bound states get embedded into the continuum can be obtained via the formulas derived in (\ref{levelcrossing0})-(\ref{levelcrossing2}) and (\ref{BIC1}), (\ref{BIC2}).

\begin{figure}
\centering
\subfloat[][$\kappa=2.9$]{\includegraphics[width=0.4\textwidth]{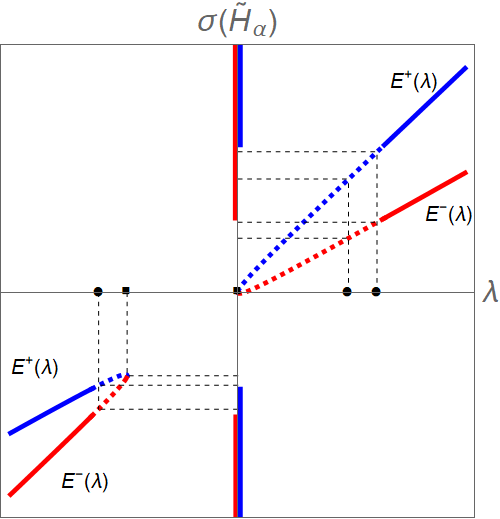}}
\hspace{2mm}
\subfloat[][$\kappa=3.5$]{\includegraphics[width=0.4\textwidth]{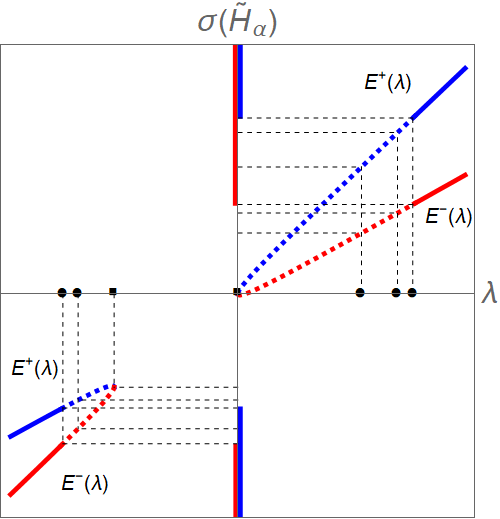}}
\caption{Eigenvalue curves $\mathbb{E}^{+}(\lambda)$ (blue) and $\mathbb{E}_{\lambda}^{-}(\lambda)$ as functions of $\lambda\in\mathbb{R}$ within the scattering eigenvalues region (solid), the non-physical eigenvalues region (dashed), and for the bound state eigenvalues (black-filled-circles). The parameters have been fixed as $U_{0}=1$, $\alpha=0.25$, and the indicated values of $\kappa$.}
\label{fig:PT-compo-2}
\end{figure}

For completeness, in Fig.~\ref{fig:PT-compo-2}, we also present the spectrum of $\mathbb{H}_{\alpha}$ as based on $\mathbb{E}^{\pm}(\lambda)$ for fixed $\alpha=0.25$, together with $\kappa=2.9$ and $\kappa=3.5$. This allows us to obtain further insight on the spectral curves for non-physical and scattering energies, as well as bound state energies. The so chosen value of $\alpha$ is such that not all the bound state energies have been embedded into the continuum.

\section{Discussion}

The idea of composite of operators presented in this paper is not restricted to the generators of the superalgebra. From the technical point of view, it is possible to make a linear combination of any set of commuting operators (e.g. of the Hamiltonian and its integrals of motion). Nevertheless, it has to be followed by providing a physical meaning of such composite operator. In our work, we showed that susy operators (\ref{susyop}) associated with the Dirac equations (\ref{r1}), (\ref{r2}) can be used quite naturally for construction of a coupled system of two Dirac fermions. The coupling is provided by the supercharge/supersymmetric transformation. The fact that both the supersymmetric Hamiltonian $\mathbb{H}_0$ and the supercharge $\mathbb{L}_1$ were differential operators of the same order allowed us to transform the kinetic term into the convenient form that suggested different Fermi velocities of the two fermions. The same approach can be applied to non-relativistic susy operators. In this context, the use of generators of the nonlinear superalgebra, where the supercharges are of the same order as the Schr\"odinger Hamiltonian, seems to be more convenient. It is also worth mentioning that the construction of composite operators is not restricted to Hermitian ones. It is worth noticing in this context that coupled system of two ($PT$-symmetric) Swanson oscillator was discussed recently in \cite{Bagchi}.

The two Hamiltonians $\mathbb{H}_0$ and $\widetilde{\mathbb{H}}_{\alpha}$ are not isospectral for $\alpha\neq0$. Nevertheless, we showed that the spectral properties of $\mathbb{H}_{\alpha}$ can be unambiguably derived from those of $\mathbb{H}_0$. Instead of isospectrality, there is spectral isomorphism between the two Hamiltonians. In this respect, the current results extend the class of isospectral systems discussed in \cite{Jakubsky}.

We observed that there are eigenstates shared by $\mathbb{H}_0$ and $\mathbb{H}_{\alpha}$ that correspond to complex eigenvalue $\lambda$ of $\mathbb{H}_0$, nevertheless, the corresponding eigenvalues $\mathbb{E}^{\pm}(\lambda)$ of $\mathbb{H}_{\alpha}$ are real. 
In general, the square-integrable solutions corresponding to the eigenvalues $\pm i$ form basis of so called deficiency subspaces of a non-self-adjoint operator. When the dimensions of the two deficiency subspaces are the same, self-adjoint extension of the operator is possible, see \cite{Gustafson} for details.
The basis vectors of the deficiency subspaces could form square-integrable solutions of (\ref{bispinor1}) with real energy. Nevertheless, there arises question whether such solutions would belong into the domain of definition of the composite Hamiltonian $\mathbb{H}_{\alpha}$. We leave analysis of this point for future work.

\section*{Acknowledgment}
V. J. was supported by GACR grant no 19-07117S. K.Z. acknowledges the support from the project “Physicists on the move II” (KINE\'O II), Czech Republic, Grant No. CZ.02.2.69/0.0/0.0/18 053/0017163; and Consejo Nacional de Ciencia y Tecnolog\'ia (CONACyT), Mexico, Grant No. A1-S-24569.

	
\end{document}